%% file: camera-ready.tex
\documentclass[acmsmall]{acmart}
\newcommand{\tool}{\textsf{LegoFuzz}\xspace}

\usepackage{graphicx} 
\usepackage{xspace}
\usepackage[frozencache=true,cachedir=minted-cache]{minted} 
\usepackage{subcaption}
\usepackage{amsthm}
\usepackage{amsmath}
\usepackage{xcolor}
\usepackage{csquotes}
\usepackage{mathtools}
\usepackage{textcomp}
\usepackage{enumitem}
\usepackage{multirow}
\usepackage{tcolorbox}
\usepackage{pifont}
\usepackage{makecell}

\usepackage{xspace}

\newcommand{\etal}{\hbox{\emph{et al.}}\xspace}
\newcommand{\eg}{\hbox{\emph{e.g.}}\xspace}
\newcommand{\ie}{\hbox{\emph{i.e.}}\xspace}

\newcommand{\etc}{\hbox{\emph{etc.}}\xspace}

\usepackage[linesnumbered,ruled,vlined]{algorithm2e}

\SetCommentSty{mycommfont}

\newcommand{\dashrule}[1][black]{%
  \color{#1}\rule[\dimexpr.5ex-.2pt]{4pt}{.4pt}\xleaders\hbox{\rule{4pt}{0pt}\rule[\dimexpr.5ex-.2pt]{4pt}{.4pt}}\hfill\kern0pt%
}

\newcommand{\para}[1]{\smallskip\noindent\textbf{#1}\xspace}
\newcommand{\tmark}{{\scriptsize\ding{228}}}

\newcommand{\circledone}[1]{{\Large\ding{172}}}
\newcommand{\circledtwo}[1]{{\normalsize\ding{173}}}
\newcommand{\circledthree}[1]{{\normalsize\ding{174}}}
\newcommand{\circledfour}[1]{{\normalsize\ding{175}}}
\newcommand{\circledfive}[1]{{\normalsize\ding{176}}}
\newcommand{\mycircled}[1]{\raisebox{.5pt}{\textcircled{\raisebox{-.9pt} {#1}}}}

\AtBeginDocument{%
  }

\setcopyright{cc}
\setcctype{by}
\acmDOI{10.1145/3763079}
\acmYear{2025}
\acmJournal{PACMPL}
\acmVolume{9}
\acmNumber{OOPSLA2}
\acmArticle{301}
\acmMonth{10}
\received{2025-03-25}
\received[accepted]{2025-08-12}

\ccsdesc[500]{Software and its engineering~Compilers}
\ccsdesc[500]{Software and its engineering~Software testing and debugging}

\keywords{Compilers, Testing, Reliability}

\begin{document}

\title[Interleaving LLMs for Compiler Testing]{Interleaving Large Language Models for Compiler Testing}

\author{Yunbo Ni}
\orcid{0009-0004-4837-6696}
\affiliation{%
  \institution{The Chinese University of Hong Kong}
  \city{Shatin}
  \country{Hong Kong}
}
\email{ybni@cse.cuhk.edu.hk}

\author{Shaohua Li}
\authornote{Corresponding author}
\orcid{0000-0001-7556-3615}
\affiliation{%
  \institution{The Chinese University of Hong Kong}
  \city{Shatin}
  \country{Hong Kong}
}
\email{shaohuali@cse.cuhk.edu.hk}

\begin{abstract}
Testing compilers with AI models, especially large language models (LLMs), has shown great promise. However, current approaches struggle with two key problems: The generated programs for testing compilers are often too simple, and extensive testing with the LLMs is computationally expensive. 
In this paper, we propose a novel compiler testing framework that decouples the testing process into two distinct phases: an offline phase and an online phase.
In the offline phase, we use LLMs to generate a collection of small but feature-rich code pieces.
In the online phase, we reuse these code pieces by strategically combining them to build high-quality and valid test programs, which are then used to test compilers.

We implement this idea in a tool, \tool, for testing C compilers. The results are striking: we found 66 bugs in GCC and LLVM, the most widely used C compilers. 
Almost half of the bugs are miscompilation bugs, which are serious and hard-to-find bugs that none of the existing LLM-based tools could find.  
We believe this efficient design opens up new possibilities for using AI models in software testing beyond just C compilers.

\end{abstract}


\maketitle

\section{Introduction}
Compilers play a fundamental role in the modern software ecosystem. However, despite significant efforts to enhance their reliability, they remain prone to bugs~\cite{chen2021compilersurvey}.  
Therefore, large-scale compiler testing is essential for identifying and eliminating these bugs. Various approaches have been proposed, such as random program generation~\cite{yang2011csmith, livinskii2020yarpgen, livinskii2023yarpgen} and mutation-based testing~\cite{le2014emi, sun2016emi}.
With the advancement of Large Language Models (LLMs), new tools have emerged to enhance compiler testing. A notable example is Fuzz4All~\cite{xia2024fuzz4all}, a universal fuzzer designed for multiple purposes, including compiler testing. It leverages LLM-generated prompts within the testing loop to iteratively generate test programs, improving the efficiency and diversity of fuzzing. 
Another tool, WhiteFox~\cite{yang2024whitefox}, is specifically designed for deep learning compilers and features a multi-agent framework. It generates prompts using relevant documentation and example code to guide LLMs in producing targeted and effective test programs.


Although current LLM-based compiler testing has demonstrated great potential, its practical deployment comes with notable challenges. In real-world applications, two major concerns arise: (1) the difficulty of ensuring the quality and validity of generated test cases and (2) the high computational cost of integrating LLMs into large-scale testing workflows.
Below, we elaborate on the two core challenges.

\begin{figure*}[tp]
    \centering
    \includegraphics[width=0.9\linewidth]{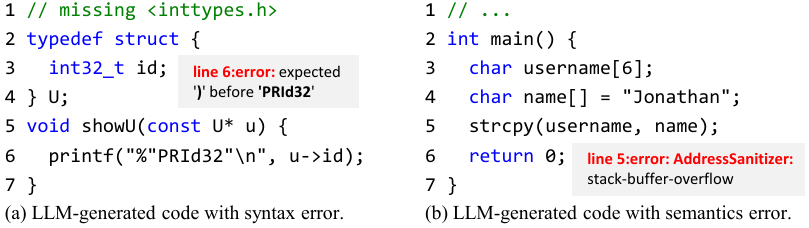}
    \vspace{-5pt}
    \caption{Examples of LLM-generated erroneous code. 
    }
    \label{fig:intro}
    \vspace{-10pt}
\end{figure*}

\para{\tmark~Challenge 1: Low quality of testing programs.} 
In compiler testing, we typically need to generate complex and valid programs, both \emph{syntactically} and \emph{semantically}.
However, the programs generated by the current LLM-based tools do not always meet these requirements.
Unlike natural languages, programs contain rich grammar and semantic constraints, which LLMs are not fully aware of~\cite{jiang2024surveylargelanguagemodels}. Thus, LLMs are fundamentally limited in that they are not guaranteed to generate valid programs. 
For example, Figure~\ref{fig:intro} (a) shows the C program generated by the model used in Fuzz4All, which contains a syntax error and cannot be compiled by GCC. Figure~\ref{fig:intro} (b) shows another LLM-generated program, which can be compiled but contains a semantic error, \ie, a buffer overflow in line 5.
Such invalid programs can only be used to find crash bugs, not miscompilation bugs.
In fact, neither Fuzz4All nor WhiteFox found any miscompilation bugs in C compilers. All GCC bugs found by Fuzz4All are related to compiler front-end crashes rather than core compiler optimizations. WhiteFox identified only two bugs in LLVM, one involving a compiler backend crash and another related to a crash in error diagnostics.
According to its own reports, Fuzz4All generates only 37.26\% valid C programs for GCC. The average length of its generated programs is just 18 lines, while WhiteFox produces slightly longer programs with an average of 21 lines based on our evaluation.
Hence, how to generate complex (\eg, thousands of lines) yet valid programs with LLMs for compiler testing is a challenging problem.

\para{\tmark~Challenge 2: High computational cost.} 
Currently, using LLMs for fuzzing is expensive, both computationally and financially. 
An effective fuzzing process is usually required to generate a large volume of test inputs in a short period~\cite{zhu2022fuzzingsurvey}. 
However, current LLM-based fuzzers are highly constrained by the sheer amount of computational cost of LLMs.
For example, Fuzz4All can generate only around 16,000 valid C programs in 24 hours. Extending these LLM-based fuzzers to produce hundreds of millions of programs would require an extremely large amount of computational or financial resources. 
In contrast, traditional fuzzers can easily achieve such volumes of data on modern hardware, as there is some code analysis checking the changes or the generated code to some degree. 
For example, Csmith generates around 1 million test programs overnight, with 99.96\% validity rate~\cite{even2023grayc, tu20221mcsmith, even2022csmithedge}. 

\para{\tmark~Our core idea.} This paper addresses the challenges mentioned above by innovatively decoupling the entire compiler testing process into two distinct phases: \emph{an offline phase} and \emph{an online phase}. 
Rather than relying solely on LLMs throughout the entire pipeline, we integrate both LLMs and traditional program generation/mutation techniques, each assigned to the phase where they are most effective.
In the offline phase, we utilize LLMs to generate small yet feature-rich and valid programs, which are then \textbf{\emph{reused as building blocks}} for the online phase. 
In the online phase, we iteratively use the building blocks to generate increasingly complex yet valid programs for compiler testing. 
This hybrid approach balances the strengths of both LLMs and traditional methods, optimizing both the quality and the cost of LLM-based compiler testing. 

\para{\tmark~Our approach: \tool.} We propose \tool, a novel compiler testing framework that strategically decouples LLM usage into two synergistic phases: \textbf{\emph{an offline phase}} for collecting high-quality code building blocks through LLM-guided transformation of real-world code, and \textbf{\emph{an online phase}} for iterative program synthesis that efficiently reuses these building blocks to generate complex test cases without further LLM invocations.
Our goal in the first offline phase is to generate programs with measurable complexity while maintaining syntactic and semantic validity. 
An observation is that LLM-generated code often exhibits recurring structural patterns that correlate with the models' pre-training data~\cite{deng2024fuzzgpt}. This observation suggests potential limitations in the diversity of generated programs, which could affect their effectiveness for compiler testing.
To address this limitation, we leverage existing open-source code as templates to guide LLMs in generating or transforming code snippets. We then validate their syntactic and semantic validity and only keep the valid ones.
All code snippets are organized as single functions to facilitate the validation as well as the future synthesis process.
Using this approach, we construct a code database containing \emph{over half a million} functions from 146 open-source projects, including system software (\eg, Linux), databases (\eg, Redis), and web servers (\eg, Nginx).


The online phase uses these functions as building blocks to synthesize larger, more complex programs. 
A crucial step in this phase is to establish dependencies between the functions. Without these dependencies, \ie, simply putting multiple functions in one program, compiling the program would be mostly equivalent to compiling each function separately, thus limiting the coherence of the generated program. 
For example, the left snippet shows two independent functions, while the right one introduces a call from \texttt{a} to \texttt{b}, forming a dependency. 
With such dependencies, we cover 3,726 more lines in LLVM and trigger more analysis and optimizations like inline and instcombine.

\vspace{5pt}
\begin{minipage}{0.3\textwidth}
    \begin{minted}[xleftmargin=1em,fontsize=\scriptsize,escapeinside=@@]{C}
int b(int y) { return y * 2; }
int a(int x) {
  if (x > 10) return 1;
  else return 0;
}
    \end{minted}
\end{minipage}
\hfill
\begin{minipage}{0.3\textwidth}
    \begin{minted}[xleftmargin=1em,fontsize=\scriptsize,escapeinside=@@]{C}
int b(int y) { return y * 2; }
int a(int x) {
  if (x > b(5)) return 1;
  else return 0;
}
    \end{minted}
\end{minipage}
\hfill
\begin{minipage}{0.3\textwidth}
\qquad\qquad\qquad\\

\end{minipage}
\vspace{5pt}

We implement two specific mechanisms to build dependencies between functions: (1) function call insertion, where we insert calls between functions based on type compatibility and semantic constraints, and (2) global variable share, where we introduce shared variables to create verifiable data dependencies between functions.

\medskip
To evaluate the effectiveness of \tool, we select two widely used modern C compilers, GCC and LLVM, as our fuzzing targets. By stress-testing their latest versions, \tool successfully uncovered 66 compiler bugs, among which 30 were miscompilation bugs, and 56 were already fixed by the developers.
In conclusion, this paper makes the following contributions:

\begin{itemize}[labelwidth=!, labelindent=0pt, itemsep=5pt, topsep=2pt]
    \item We present \tool, an LLM-based compiler testing framework that decouples the program generation process into offline and online phases, enabling the effective generation of high-quality test programs with LLMs.

    \item We design a novel real-world code-aligned prompting method to guide LLMs in generating code snippets with a diverse range of features.
    
    \item We propose a novel iterative program synthesis method that strategically combines multiple code snippets to complex, feature-rich yet semantically valid programs.

    \item We implement \tool for C compiler testing and construct a comprehensive database consisting of over half a million functions generated by LLMs. We conduct an extensive fuzzing campaign with \tool, discovering 66 unique bugs in widely used modern C compilers, \ie, GCC and LLVM.
\end{itemize}

\medskip
\tool has been open-sourced at \url{https://github.com/cuhk-s3/LegoFuzz}. We believe that this fuzzing framework offers a promising direction for applying LLMs to real-world compiler testing scenarios. 
Its modular and extensible design philosophy allows easy adaptation to other systems under test, broadening its applicability beyond C compiler testing. 

\begin{figure*}[tp]
    \centering
    \includegraphics[width=0.97\linewidth]{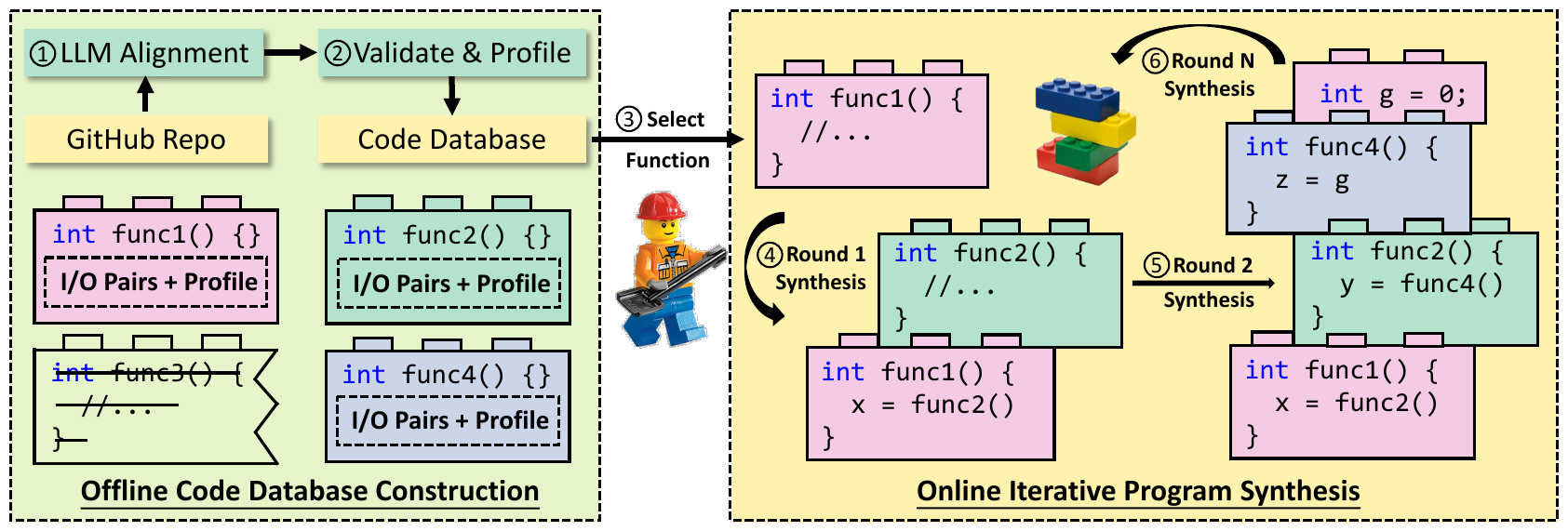}
    \caption{Design overview of \tool. The left part is the offline code database construction phase, which provides the building blocks for the online iterative program synthesis on the right.}
    \label{fig:overview}
    \vspace{-10pt}
\end{figure*}

\section{Design Overview}\label{sec:design}

This section outlines the high-level design of our proposed framework \tool.
The core idea of \tool is to separate the whole testing process into offline and online phases. Figure~\ref{fig:overview} shows the high-level workflow of \tool. 
The offline phase queries an LLM to collect valid code snippets, which enables us to control the quality of code as well as the cost of LLM invocations. 
In contrast, the online phase eliminates the dependency on LLMs by \textbf{reusing} these pre-generated code snippets. Through our proposed \textit{iterative program synthesis}, the online phase constructs increasingly complex yet valid programs, which can be used to test compilers. 
Below, we provide a more detailed overview of our \tool:

\begin{enumerate}[labelwidth=!, labelindent=0pt, itemsep=5pt, topsep=5pt]
    \item \textbf{Phase 1: Offline Code Database Construction with LLMs.} Firstly, we utilize an LLM to generate a diverse set of code snippets. 
    At a high level, we propose to use real-world code snippets from GitHub repositories as the guiding templates to enable LLMs to generate diverse code snippets. 
    As illustrated in the left part of Figure~\ref{fig:overview}, the first step is to collect code via LLM-based alignment. In this example, four functions, \ie, \texttt{func1}, \texttt{func2}, \texttt{func3}, and \texttt{func4}, are collected.
    In the second step, we analyze each function to determine its validity and get its run-time profiling information. Invalid functions with syntactic or semantic errors are discarded. In this example, the function \texttt{func3} is excluded.
    All collected code snippets, along with their profiling information, are stored in the code database.

    \item \textbf{Phase 2: Online Iterative Program Synthesis.}
    Although the code produced by the LLM contains diverse features, they are still simple, \eg, averagely 30 lines, and we could not find any interesting compiler bugs according to our evaluation. 
    In this paper, we propose an iterative program synthesis method to generate complex programs with these LLM-generated code snippets as building blocks.
    For example, in the right part of Figure~\ref{fig:overview}, we select the function \texttt{func1} as the seed function in Step \mycircled{3}. During Round 1 of synthesis in Step \mycircled{4}, \texttt{func1} is modified by replacing the value of \texttt{x} with a call to \texttt{func2()}. \tool ensures that this replacement introduces no side effects by leveraging the pre-collected profiling results. 
    Similarly, in the following two iterations of synthesis in Step \mycircled{5}, a call to \texttt{func4} is inserted into \texttt{func2}, and a read from the global variable \texttt{g} is inserted into \texttt{func4}.
    The example program can undergo multiple synthesis iterations in Step \mycircled{6}, allowing for the \textbf{reuse} of additional building blocks. 
    After synthesis, we generate a driver main function that calls all used functions such as \texttt{func1}, \texttt{func2}, and \texttt{func4}, and operates on global variables like \texttt{g} to perform mutations and print checksums for fuzzing. We detail this process in Section~\ref{sec: impl}.
    Through this \textit{iterative program synthesis} process, the final synthesized program can grow significantly in complexity, ultimately reaching \textbf{tens of thousands of lines} in length. 
    Because this online phase directly uses the established code database, it can quickly produce a large number of complex programs without querying LLMs. 
    In some sense, we are maximizing the utility of LLMs' outputs with our iterative synthesis while eliminating the need for continual LLM usage to significantly reduce costs. 
\end{enumerate}

\tool aims at efficiently and effectively generating complex yet valid programs using LLMs. This approach enables the seamless integration of LLM-based fuzzing into real-world testing pipelines. 
The framework's modular architecture provides great flexibility, allowing developers to easily adapt and extend its capabilities.

\begin{figure}[!tp]
    \centering
    \begin{minipage}{0.4\linewidth}
        \centering
        \includegraphics[width=0.9\linewidth]{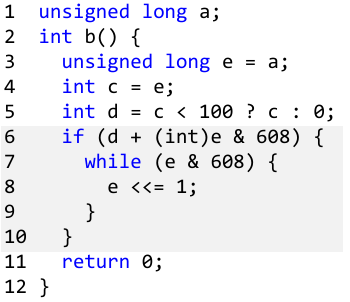}
        \caption{Code that triggers a crash in GCC.}
        \label{fig:crash-gcc}
    \end{minipage}
    \hfill
    \begin{minipage}{0.55\linewidth}
        \centering
        \includegraphics[width=0.9\linewidth]{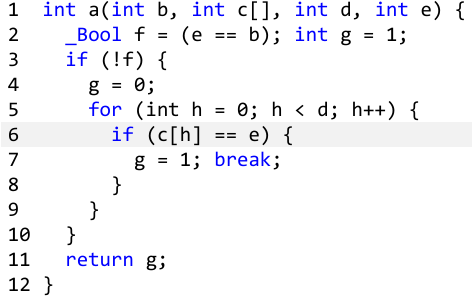}
        \caption{An LLM-generated program with potential overflow.}
        \label{fig:fastsocket-transformed}
    \end{minipage}
    \vspace{-10pt}
\end{figure}

\section{\tool}\label{sec:design}

In this section, we describe the technical details of our \tool framework. 
Section~\ref{subsec:offline} details the offline code collection with large language models, while Section~\ref{subsec:online} introduces the online iterative program synthesis.

\subsection{Offline Code Database Construction with Large Language Models}\label{subsec:offline}

The goal of the offline phase is to collect a set of code snippets using LLMs, which will serve as the building blocks for the later online phase.
We use an LLM to first generate these code snippets. Then, we validate them and carefully analyze how they work.
By the end of this offline phase, we will have a code database $\mathcal{D}_F$, where each entry $E_i \in \mathcal{D}_F$ contains two essential parts, \ie, $E_i=\{F_i, prof_i\}$: 
(1) $F_i$ is a code snippet existing as a function, and (2) $prof_i$ is a detailed profile that records how the function $F_i$ behaves when run, such as variable types and variable values.
Specifically, each $F_i$ meets the following two criteria:
\begin{itemize}[labelwidth=!, labelindent=10pt, itemsep=5pt, topsep=5pt]
    \item \textbf{Expressive:} Compilers must handle a wide variety of program structures and features. For effective compiler testing, our generated programs should contain complex and expressive elements such as non-trivial control flows and intricate program semantics.
    Figure~\ref{fig:crash-gcc} shows one of our reported crash bugs in GCC. This bug is triggered by the \emph{complex, nested, and unbounded loop structure} in lines 6–10. 
    Such sophisticated code patterns are difficult for LLMs to generate through standard natural language prompting. 
    To overcome this limitation, we introduce real-world code-aligned prompting, a technique that uses real-world code as templates to guide LLM code generation. We explain this approach in detail in Section~\ref{subsubsec:real-align}.
    
    \item \textbf{Valid:} All generated code must be both syntactically and semantically valid, adhering to the language specification. 
    For C programs, this means passing compiler grammar checks and containing no undefined behaviors at runtime. 
    This property is crucial because \emph{compilers are only designed to correctly compile valid code}~\cite{online:ub}, making it impossible to detect genuine miscompilation bugs using invalid code.
    In our design, we ensure validity through rigorous validation and profiling (see details in Section~\ref{subsubsec:database}).
    Figure~\ref{fig:fastsocket-transformed} illustrates a function transformed by an LLM from the open-sourced \texttt{Fastsocket}~\cite{online:fastsocket} project. This function is only valid when the input variable \texttt{d} is less than the length of array \texttt{c[]}. For instance, when \texttt{d}=1 and \texttt{len(c)}=2, the function works correctly. However, if \texttt{d}=3, a stack buffer-overflow occurs at line 6, making the function invalid.
    To prevent such issues, for each function, we validate it with diverse randomized inputs and only keep inputs that exhibit no undefined behaviors in the function. With these inputs, we ensure each function can be safely used in our online program synthesis phase.
    We detail this process in Section~\ref{subsubsec:database}.
\end{itemize}


\begin{figure}[t]
\centering
\includegraphics[width=0.97\linewidth]{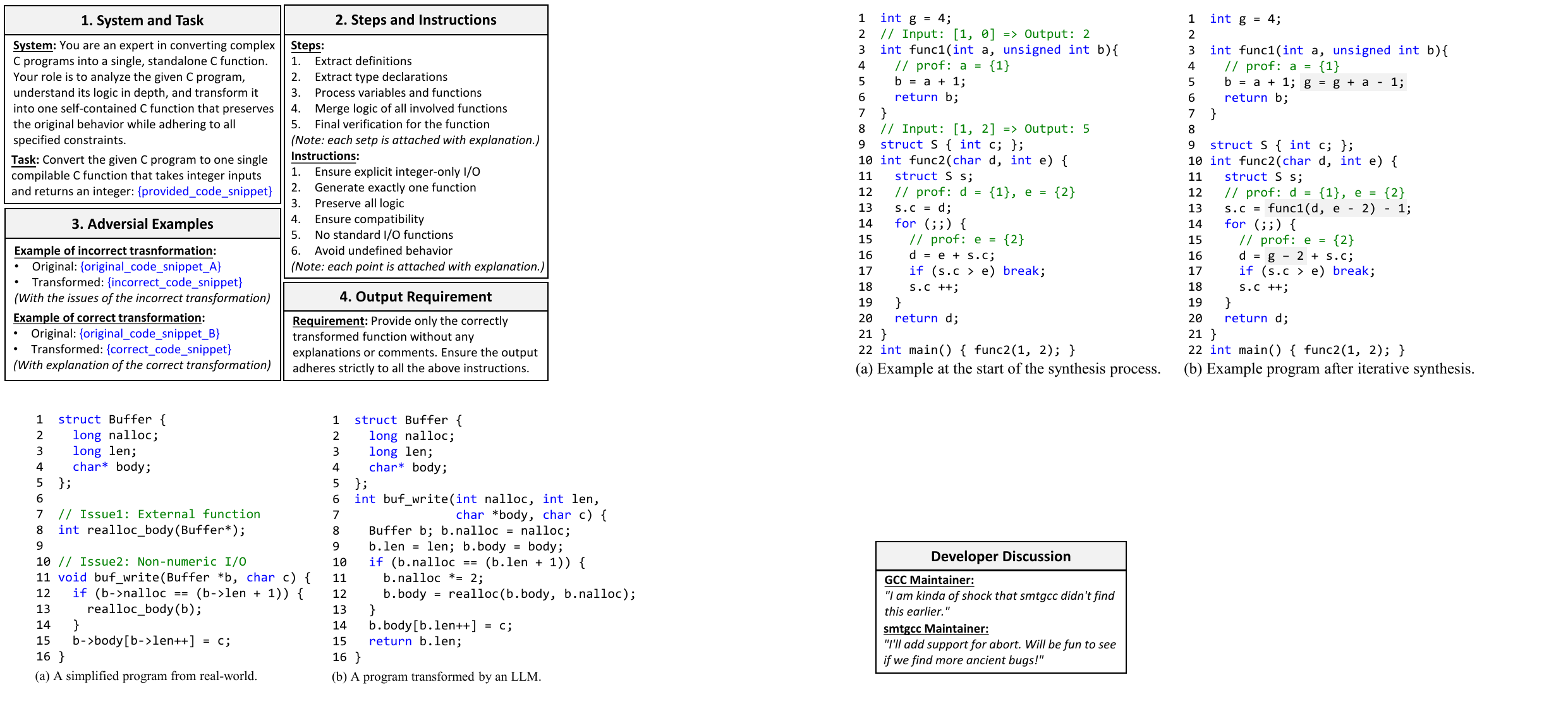}
\vspace{-10pt}
\caption{
The program in (a) is from an open-source project. LLMs transform it to the function in (b).
}
\label{fig:alignment}
\vspace{-10pt}
\end{figure}

\subsubsection{Real-world Code-aligned Code Generation}\label{subsubsec:real-align}

~\vspace{5pt}

LLMs have shown a superior ability in code generation, but it remains challenging to \emph{instruct} LLMs to produce \emph{a large volume of feature-rich code} suitable for compiler testing.
Take the code in Figure~\ref{fig:alignment} (a) as an example. It contains a non-primitive struct and manipulates a struct variable inside the \texttt{buf\_write} function.
\emph{How can we instruct LLMs to generate such code?}
Precisely describing such code features in natural language is hard, let alone the automated generation of millions of such code.
Fortunately, developers have already written such code in real-world projects. 
The large volume of open-sourced projects exercise a wide range of language features and code patterns, such as pointer manipulation, byte-level operations, non-trivial control flows, \etc
Our idea is to use these real-world codes as the template to guide LLMs.
In fact, the code in Figure~\ref{fig:alignment} (a) is from an open-source project \texttt{8cc}\cite{online:8cc}, and the code in Figure~\ref{fig:alignment} (b) is the transformed version by an LLM that can be used in our \tool framework.
Below, we detail two critical designs of our real-world code-aligned code generation approach.

\para{\tmark~Where and how to get real-world code snippets?} 
In our framework, real-world code snippets can be sourced from a variety of origins with varying levels of granularity. These snippets can range from entire programs to individual functions, as LLMs possess a fundamental ability to understand the surrounding context of a given snippet.
In order to control the length of our prompts and facilitate the profiling process, we extract functions from open-source projects as well as the necessary context, such as type definitions and global variables.
These real-world functions will then be used as the template to guide LLMs.
We will justify our choice of having functions as the granularity in Section~\ref{sec:discussion}.

\para{\tmark~How to prompt LLMs to generate real-world code-aligned code?} 
To preserve the expressiveness of LLM-generated code, we introduce \textit{real-world code-aligned prompting}, a technique that guides LLMs in transforming real-world code snippets. 
To ensure proper transformation, we specifically design the prompt with instructions that enforce two levels of alignment: 

\begin{itemize}[labelwidth=!, labelindent=10pt, itemsep=5pt, topsep=5pt]
    \item \textbf{Syntax-level alignment.} To facilitate the input generation of later profiling process, \tool requires the transformed function (1) to be numeric, \ie, both its input and output have only numeric types or pointers to numeric types, and (2) to be the sole function with no additional function definitions present. 
    For example, the transformed function in Figure~\ref{fig:alignment} (b) satisfies this requirement.
    This requirement ensures that we can easily model a function's semantics with its numeric input and output, which eases our implementation of both the later profiling and online program synthesis.
    \item \textbf{Semantics-level alignment.} One might think that \tool cannot transform real-world code having complex types (\eg, \texttt{struct Buffer} in Figure~\ref{fig:alignment} (a)) or multiple functions (\eg, \texttt{realloc\_body} and \texttt{buf\_write} in Figure~\ref{fig:alignment} (a)). 
    However, we guide the LLMs to preserve the full semantics of the original program. 
    For complex types like \texttt{struct}, initialization can be deferred to the function body. 
    For programs with multiple functions, the transformed output aims to inline all logic into a single function while maintaining the original program's behavior.
    For example, in Figure~\ref{fig:alignment} (b), the \texttt{struct} definition is moved inside the function body, and the struct input is replaced with its numeric fields. 
    Furthermore, LLMs can deduce that \texttt{realloc\_body} likely reallocates memory for the \texttt{body} field of \texttt{Buffer}. Consequently, it can be directly implemented using the standard C memory management function \texttt{realloc}. This inference achieves the semantics-level alignment with the original program. 
\end{itemize}

Since LLMs do not always follow users' instructions, even if we instruct LLMs to satisfy the above alignments, the transformed function may still violate the above requirements.
Thus, we validate and filter these transformed functions in the next step.

\begin{figure*}[tp]
    \centering
    \includegraphics[width=0.93\linewidth]{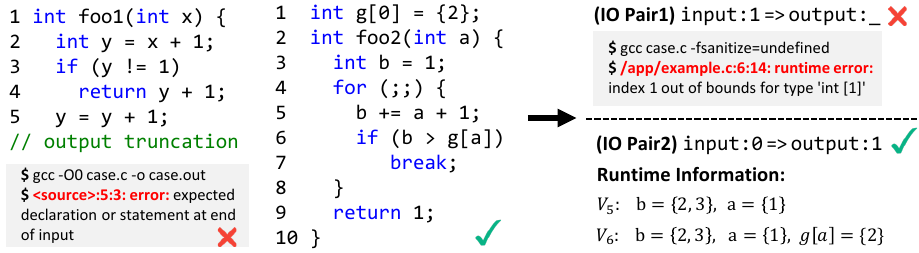}
    \vspace{-5pt}
    \caption{Process of code database construction.}
    \label{fig:profile}
    \vspace{-10pt}
\end{figure*}

\subsubsection{Code Database Construction}\label{subsubsec:database}

~\vspace{5pt}

In this part, we describe how we validate the transformed functions and profile the valid ones.
The profiling process is inspired by Hermes~\cite{sun2016emi} and Creal~\cite{creal2024li}.

Although we have instructed the LLM to align the generated functions with the original programs, the output may still exhibit several issues that impact their validity. For instance, current LLMs are constrained by the \texttt{max\_completion\_tokens} parameter~\cite{online:max-completion-tokens}, which limits the number of output tokens. As a result, generated programs are prone to truncation, leaving them incomplete and cannot be compiled. 
For example, the first function \texttt{foo1} in Figure~\ref{fig:profile} is incomplete due to truncation, resulting in a compilation error. This function is thus excluded from the code database.

For syntactically valid functions, it is crucial to ensure their runtime behaviors are also correct.
Since runtime behaviors are associated with input, different inputs may lead to different behaviors.
For instance, the second function \texttt{foo2} in Figure~\ref{fig:profile} is syntactically valid, but it may lead to runtime errors when the input \texttt{a} is greater than the length of the global array \texttt{g[]}.
To ensure runtime validity, we construct a main program that invokes the candidate function with a randomly generated input based on its input types, and then use multiple sanitizers~\cite{kosta2016sanitizer1, kosta2016sanitizer2}—including ASan, UBSan, MSan, and TypeSan—together with CompCert~\cite{online:compcert} to cross check if the input triggers runtime errors, a common practice in checking the runtime validity of a program~\cite{le2014emi,creal2024li}. 
Since sanitizers can miss certain undefined behaviors in theory~\cite{li2024ubfuzz}, we did not encounter such cases during our evaluation. 
We believe that this is due to our combined use of multiple tools, which makes it highly unlikely that false alarms will occur.
If the input does not trigger runtime errors, we then use this input to profile the function and collect the intermediate runtime information.
The top right part of Figure~\ref{fig:profile} shows the first generated input (\ie, \texttt{1}) for the function \texttt{foo2}, which triggers a stack buffer-overflow. Thus, we discard this input and proceed to generate another input.
The second input (\ie, \texttt{0}) is valid, so we use it to profile the function and collect its runtime profile $prof$.
For each valid function $F_i$, its profile $prof_i$ contains the following information:
\begin{itemize}[labelwidth=!, labelindent=10pt, itemsep=3pt, topsep=2pt]
    \item \textbf{input:} The input value of the function, such as \texttt{0}.
    \item \textbf{output: } The output of executing the function with the input, such as \texttt{1}.
    \item \textbf{expression values at each line:} The values of expressions evaluated at each line of the function, such as the values of \texttt{b}, \texttt{a}, and \texttt{g[a]} at lines 5 and 6, as shown in the bottom right part of Figure~\ref{fig:profile}.
    We focus on basic expressions that hold values, including variables, array accesses, pointer dereferences, and field accesses.
\end{itemize}

In practice, \tool can generate multiple valid inputs for one function, which means one $F_i$ may have multiple profiles $prof_{i}$. 
These functions, together with their profiles, are stored in the code database $\mathcal{D}_F$:
 (1) each function $F_i$ is syntactically valid,
 and (2) invoking $F_i$ with any input from the corresponding $prof_{i}$ is semantically valid.

A natural question arises: \textbf{\emph{can these individual functions alone detect compiler bugs?}}
Our evaluation in Section~\ref{subsec: rq1} shows that the answer is negative.
This finding directly motivates our approach to synthesize complex test programs by combining multiple functions with rich inter-dependencies rather than relying on individual functions alone.

\begin{figure}[tp]
\centering
\includegraphics[width=\linewidth]{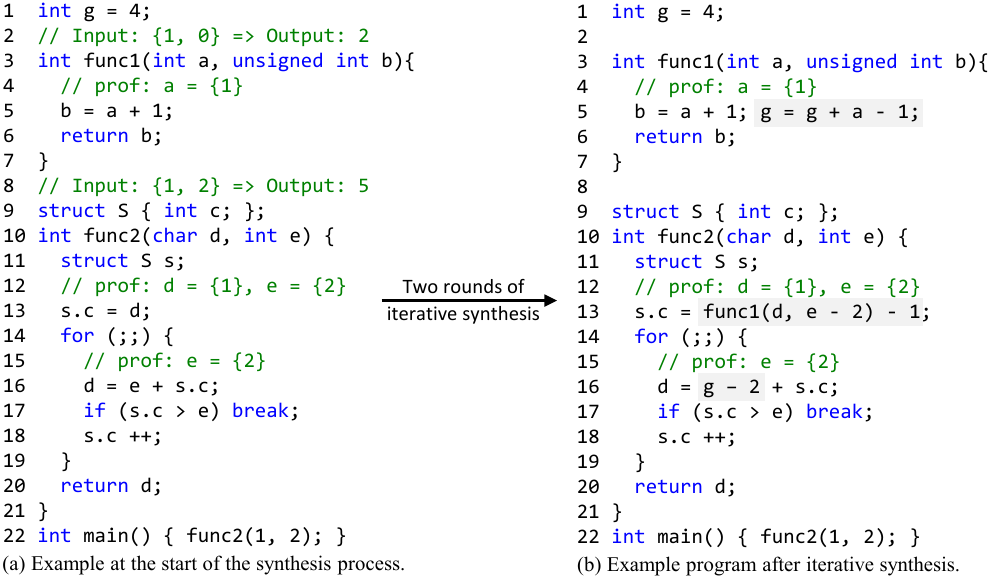}
\vspace{-10pt}
\caption{
The example in (a) is the beginning of one synthesis iteration. One possible synthesis result after two iterations is shown in (b). The synthesized part is highlighted in \colorbox{gray!10}{gray} in (b).
}
\label{fig:syn}
\end{figure}

\subsection{Online Iterative Program Synthesis}\label{subsec:online}

As has been discussed before, effective compiler testing often requires complex programs with rich features. Given the code database, our target is to synthesize complex yet valid programs. 
One may argue that we can simply put multiple functions into one file to get a complex program, as shown in Figure~\ref{fig:syn} (a), where we put two individual functions, \texttt{func1} and \texttt{func2}, into one file.
However, since there are no dependencies between these two functions, compilers will deal with them separately at the module level. 
Thus, it is almost equivalent to testing the compilers with each function individually. 



Our core idea is to couple multiple functions by building \textbf{complex dependencies} among them. 
We build the dependencies in two ways: (1) function call insertion and (2) global variable share.
For example, the program in Figure~\ref{fig:syn} (b) shows the resulting program from Figure~\ref{fig:syn} (a) by building dependencies with \tool.
In Figure~\ref{fig:syn} (b) at line 13, we replace the expression \texttt{d} with a call to \texttt{func1} to build dependency; The global variable \texttt{g} is written in \texttt{func1} at line 5 and read by \texttt{func2} at line 16, further building dependencies between them.

\begin{algorithm}[tp]
\caption{Iterative Program Synthesis}
\label{alg:generate}
\small
\SetFuncSty{textsf}
\DontPrintSemicolon
\SetKwInput{KwInput}{Input}                
\SetKwInput{KwOutput}{Output}              
\SetKwFunction{random}{FlipCoin}
\SetKwFunction{Driver}{SynthDriverProgram}
\SetKwFunction{SelectFunction}{SelectFunction}
\SetKwFunction{GenerateGlobalVars}{GenerateGlobalVars}
\SetKwFunction{GetProfile}{GetProfile}
\SetKwFunction{GetMatchedExpr}{GetMatchedExpr}
\SetKwFunction{random}{FlipCoin}
\SetKwFunction{IsAlive}{IsAlive}
\SetKwFunction{GetFunction}{GetFunction}
\SetKwFunction{GetVar}{GetGlobalVar}
\SetKwFunction{SynFuncCall}{SynFuncCall}
\SetKwFunction{InsertFunc}{InsertFunc}
\SetKwFunction{InsertFuncDecl}{InsertFuncDecl}
\SetKwFunction{SynGlobal}{SynGlobal}
\SetKwFunction{InsertGlobal}{InsertGlobalVar}

\SetKwFunction{FSynth}{Synthesis}

\SetKwProg{Proc}{procedure}{:}{\KwRet}
\Proc{\FSynth{Code Database $\mathcal{D}_F$, Iteration Number $\mathcal{N}$}}{
    $\mathcal{G} \gets \GenerateGlobalVars()$\;
    
    $Seed \gets \SelectFunction(\mathcal{D}_F)$\;
    $\mathcal{P} \gets \Driver(Seed)$\;

    $F\_list \gets [Seed]$\;

    \Repeat{$\mathcal{N}$ times}{
        $Target \gets \GetFunction(F\_list)$\;
        $\mathit{prof} \gets \GetProfile(Target)$\;
        
        $\mathcal{E} \gets \GetMatchedExpr(\mathit{prof})$\;
        \ForEach{$expr \in \mathcal{E}$}{
            \tcp{\small{randomly decide if to sythesize $expr$}}
            \If{
            $\random()$}{
                \tcp{\small{randomly choose \textit{function call insertion} or \textit{global variable share}}}
                \If{$\random()$}{
                    $F \gets \SelectFunction(\mathcal{D}_F)$\;
                    \If{$F \notin F\_list$}{
                        $expr' \gets \SynFuncCall(expr, F, \mathit{prof})$\;
                        $\mathcal{P} \gets \InsertFunc(\mathcal{P}, expr, expr', Target)$\;
                        $F\_list$\texttt{.append($F$)}\;
                    }
                }
                \Else{
                    $\texttt{g} \gets \GetVar(\mathcal{G})$\;
                    $expr' \gets \SynGlobal(expr, \texttt{g}, \mathit{prof})$\;
                    $\mathcal{P} \gets \InsertGlobal(\mathcal{P}, expr, expr', Target)$\;
                }
            }
        }
    }

    \Return{$\mathcal{P}$}\;
}
\end{algorithm}

Algorithm~\ref{alg:generate} provides the algorithmic sketch of our iterative program synthesis approach. 
We use the example in Figure~\ref{fig:syn} to illustrate the synthesis process. 
Given a code database $\mathcal{D}_F$ and a predefined iteration number $\mathcal{N}$, our generator operates as follows: 

\begin{enumerate}[label={\textbf{Step~\arabic*.}}, labelwidth=!, labelindent=30pt, itemsep=5pt, topsep=2pt]
    \item \emph{Generate Global Variables} (line 2): Similar to Csmith and other generative black-box tools~\cite{yang2011csmith, livinskii2020yarpgen, livinskii2023yarpgen}, we first generate a set of global variables $\mathcal{G}$ with random numeric types and values. For example, the global variable \texttt{g} in Figure~\ref{fig:syn} (b) is generated with a random value of $4$.

    \item \emph{Prepare Seed Program} (lines 3-5): A function is randomly selected from $\mathcal{D}_F$ as the seed function. Next, we generate the driver program $\mathcal{P}$ to invoke it with an input. The seed function is then added to the list of used functions $F\_list$. 
    For example, \texttt{func2()} is the selected seed function in Figure~\ref{fig:syn} (b), and we synthesized the driver main function to invoke it with an input selected from its profile, \ie, \{1, 2\}.

    \item \emph{Select Target Function} (lines 7-9): For each iteration, the process begins by selecting a target function, denoted as $Target$, which will be used in the future synthesis. A set of matched expressions $\mathcal{E}$ is then extracted from its stored profiling result $\mathit{prof}$.  
    Since there is only one function in the first iteration, \ie, \texttt{func2()} in Figure~\ref{fig:syn} (a), the target function is \texttt{func2()}. The matched expressions are \texttt{d} at line 13, \texttt{e} and \texttt{s.c} at lines 16 and 17.
    The high-level guideline of selecting these matched expressions is to choose the ones that can be replaced by other expressions with the same runtime values. Note that the runtime values of these expressions under the input are available and stored in $\mathit{prof}$. For example, the runtime value of \texttt{d} is $1$ at line 13, as annotated in line 12 in Figure~\ref{fig:syn} (a).

    \item \emph{Build Dependency by Function Call Insertion} (lines 13-17): If the expression is selected to be synthesized with a function call, the generator first selects a function $F$ from $\mathcal{D}_f$. To avoid potential stack overflow from recursive or cyclic calls, the generator checks if $F$ is already in the $F\_list$, \ie, whether $F$ has been used or not. If $F$ is not in $F\_list$, a new expression $expr'$ with a call to $F$ is synthesized to replace the original expression $expr$, and $F$ is added to $F\_list$. 
    For example, from Figure~\ref{fig:syn} (a) to (b), the expression \texttt{d} at line 13 is replaced by a function call to \texttt{func1()}.
    We will discuss the details of function call insertion in Section~\ref{subsubsec:syn-call}.

    \item \emph{Build Dependency by Global Variable Share} (lines 19-21): If the expression is selected to be synthesized with a global variable, the generator randomly selects a global variable $g$ from $\mathcal{G}$. It then synthesizes a read or write operation for $g$ with $expr$.
    For example, from Figure~\ref{fig:syn} (a) to (b), the expression \texttt{e} is replaced by \texttt{g - 2} in line 16. Another usage of the global variable \texttt{g} is also shown at line 5, where we write \texttt{g} with \texttt{a} in the next iteration. 
    The read or write from/to the same global variable builds dependencies between functions \texttt{func1()} and \texttt{func2()}.
    We will discuss the details of global variable share in Section~\ref{subsubsec:syn-global}.
\end{enumerate}

As shown in the algorithm, the synthesis iterates over all matched expressions, indicating that multiple insertions can occur in a single iteration. Consequently, the finally synthesized program may contain more than one function call. 
Additionally, by using a predefined iteration number $\mathcal{N}$, the size of the synthesized program remains controllable.
The details of building dependencies in Steps 5 and 6 are discussed in Section~\ref{subsubsec:syn-call} and Section~\ref{subsubsec:syn-global}. 
The key to building dependencies is to ensure that the synthesized program has the same semantics as the seed program. For example, the final synthesized program in Figure~\ref{fig:syn} (b) has the same output as the seed program in Figure~\ref{fig:syn} (a).
This semantic preservation relies on the valid profiling results obtained in the offline phase, which guide the safe dependency construction.
Since all profiling is completed offline, it does not impact the efficiency of the online synthesis process. 

\begin{algorithm}[tp]
\caption{Function Call Insertion.}
\label{alg:synfunc}
\small
\SetFuncSty{textsf}
\DontPrintSemicolon
\SetKwInput{KwInput}{Input}                
\SetKwInput{KwOutput}{Output}              

\SetKwFunction{FGen}{SynFuncCall}
\SetKwFunction{synthesis}{SynthesizeExpression}
\SetKwFunction{getstable}{GetStableVariables}
\SetKwFunction{isstable}{IsStable}
\SetKwFunction{getval}{GetValue}

\SetKwProg{Proc}{procedure}{:}{\KwRet}
\Proc{\FGen{Target Expression $expr$, Function $F$, Profile $\mathit{prof}$}}
{
    \texttt{[}$\overline{\texttt{inp}_1},$ $\overline{\texttt{inp}_2},$ $\cdots,$ $\overline{\texttt{inp}_m}$ \texttt{]} $\gets$ $F$\texttt{.input}\;
    \texttt{FC} $=$ `` $F_i$.\texttt{name} \texttt{(<para>$_1$, <para>$_2$, $\cdots$, <para>$_m$)}''\;

    \ForEach{$k \in [1 \dots m]$}{
        $V \gets $ $\getstable(\mathit{prof})$\;
        $para$ $\gets$ \synthesis$(V, \ \overline{\texttt{inp}_k})$ \;
        $\texttt{FC}$\texttt{.Substitute}\texttt{(<para>$_k$, para)}\;
    }

    \If{$\isstable(expr)$}{
        $val \gets \getval(expr)$\;
        $expr' \gets \synthesis(\texttt{FC}, val)$\; 
    }
    \Else{
        $expr' \gets expr + \synthesis(\texttt{FC}, 0)$\;
    }
    
    \Return{$expr'$}\;
}
\end{algorithm}

\subsubsection{Build Dependency by Function Call Insertion}\label{subsubsec:syn-call}

~\vspace{5pt}

Algorithm~\ref{alg:synfunc} outlines the procedure for synthesizing an expression using a function call. 
It begins by extracting the inputs of $F$ stored in the code database and constructs a function call with parameter placeholders that match the number of inputs (lines 2-3). 
For each placeholder, the set of stable variables $V$—those that exhibit only one run-time value at a given location—is collected from the stored profiling results $\mathit{prof}$ (line 5).
The placeholder is then substituted with a synthesized parameter $para$ that aligns with the value (lines 6-7). 
The expression $expr$ is then handled based on its stability in the following two cases: 
(1) $expr$ is stable (lines 8-10): The corresponding stable value $val$ is retrieved, and a new expression $expr'$ is synthesized by combining the return value of $F$ with $val$. 
(2) $expr$ is unstable (lines 11-12): A synthesized expression synthesized with $\textit{FC}$ equal to $0$ is concatenated with the original expression to form the new expression $expr'$.

\para{Example.} 
As shown in Figure~\ref{fig:syn}(a), the target expression $d$ in line 13 is stable with a single value of $1$. Our target is to synthesize a new expression with a call to \texttt{func1()} that also evaluates to 1:

\begin{itemize}[labelwidth=!, labelindent=5pt, itemsep=3pt, topsep=2pt, leftmargin=15pt]
    \item (Line 2 in Algorithm~\ref{alg:synfunc}) We extract the input list for \texttt{func1()} as $\{\texttt{1}, \texttt{0}\}$.
    \item (Line 3 in Algorithm~\ref{alg:synfunc}) We construct the function call \texttt{FC} as ``\texttt{func1(<para>$_1$, <para>$_2$)}''.
    \item (Lines 5-6 in Algorithm~\ref{alg:synfunc}) We first retrieve the set of stable variables $V = \{d, e\}$, all of which have only one runtime value. For each input, we generate \texttt{para$_1$} as the expression $d$, which evaluates to \texttt{1}, and \texttt{para$_2$} as $e - 2$, which evaluates to \texttt{0}. Therefore, \texttt{FC} is updated with the new parameters, resulting in ``\texttt{func1($d$, $e - 2$)}'', which is semantically equivalent to ``\texttt{func1(1, 0)}''.
    \item (Lines 8-10 in Algorithm~\ref{alg:synfunc}) Since the target expression $d$ is stable, we retrieve its value, \ie, $val = 1$. Since the return value of \texttt{func1()} is \texttt{2}, we then synthesize the new expression $expr'$ as \texttt{func1($d$, $e - 2$)} $ - 1$, which evaluates to \texttt{1} and equals to the runtime value of the target expression $d$.
    The ``-1'' after the function call is to massage the return value of \texttt{func1()} to the same value as the target expression $d$. In our implementation, we used several operators, such as ``+'' and ``-'', to achieve this.
\end{itemize}

\subsubsection{Build Dependency by Global Variable Share}\label{subsubsec:syn-global}

~\vspace{5pt}

Algorithm~\ref{alg:synglobal} describes the process of synthesizing global variable share. It breaks down into synthesizing read and write operations to the global variables. 
The construction of a read expression is similar to a function call insertion. Specifically, the new expression $expr'$ is synthesized based on the stability of the target expression $expr$. If $expr$ is stable, the stable value $val$ is used to construct a new expression $expr'$ with the same value (lines 4-5). For unstable expressions, we synthesize a new expression that equals 0 and concatenate it with $expr$ (lines 9-10).  
When synthesizing a write operation, we need to ensure that the generated expression does not alter the program semantics unexpectedly. 
The reason is that global variables are shared by multiple functions, and their values need to be statically known so that our generator can precisely control the program semantics.
We achieve this by only using stable variables for global variable writes.
A write statement is constructed by combining $g$ and the expression $expr - val$, which is 0 as $val$ is the runtime value of $expr$. This ensures that the synthesized write operation does not change the runtime value of $g$ (line 7).
This write operation is then inserted into the program at an appropriate location (line 8). 

\begin{algorithm}[tp]
\caption{Global Variable Share.}
\label{alg:synglobal}
\small
\SetFuncSty{textsf}
\DontPrintSemicolon
\SetKwInput{KwInput}{Input}                
\SetKwInput{KwOutput}{Output}              

\SetKwFunction{FGen}{SynGlobal}
\SetKwFunction{random}{FlipCoin}
\SetKwFunction{synthesis}{SynthesizeExpression}
\SetKwFunction{getstable}{GetStableVariables}
\SetKwFunction{isstable}{IsStable}
\SetKwFunction{getval}{GetValue}
\SetKwFunction{insertafter}{InsertStatementAfter}

\SetKwProg{Proc}{procedure}{:}{\KwRet}
\Proc{\FGen{Target Expression $expr$, Global Variable $G$, Profile $\mathit{prof}$}}
{
    \If{$\isstable(expr)$}{
        $val \gets \getval(expr)$\;
        \If{$\random()$}{
        $expr' \gets \synthesis(G, val)$\; 
        }
        \Else{
        $stmt\_write =$ ``$G = G + (expr - val);$''\;
        $\insertafter(stmt\_wirte)$\;
        }
    }
    \Else{
        $expr' \gets expr + \synthesis(G, 0)$\;
    }
    
    \Return{$expr'$}\;
}
\end{algorithm}

\para{Example.}
We show how to synthesize the read and write operations for the global variable $g$ from Figure~\ref{fig:syn} (a) to (b) with Algorithm~\ref{alg:synglobal}:

\begin{itemize}[labelwidth=!, labelindent=5pt, itemsep=3pt, topsep=2pt, leftmargin=15pt]
    \item (Lines 2-3 in Algorithm~\ref{alg:synglobal}) Suppose the target function is \texttt{func2()} and the target expression is $e$ at line 16 in the first iteration. Since $e$ is stable, we can generate either read or write operations for the global variable $g$. We retrieve the value of $e$ and assign $val$ as $2$.
    \item (Lines 4-5 in Algorithm~\ref{alg:synglobal}) We synthesize the new expression $expr'$ as $g - 2$, which evaluates to the same value $2$ as the original expression $e$. Then, we replace the original expression $e$ with $expr'$ in the program.
    \item (Line 6-8 in Algorithm~\ref{alg:synglobal}) In the next iteration, suppose we select \texttt{func1()} as the target function and $a$ at line 5 as the target expression, we generate the write statement ``$g = g + a - 1;$'' and insert it after this location. Because $a-1$ is $0$, the write operation does not change the runtime value of $g$.
\end{itemize}

\subsection{Beyond Creal: Design Innovations in \tool}

\tool not only addresses the fundamental challenges faced by existing LLM-based compiler testing tools, but also introduces key innovations that go beyond the capabilities of traditional fuzzing frameworks.
Creal~\cite{creal2024li} serves as a representative approach in this space, combining real-world code with Csmith-generated seeds to construct test programs. 
Thus, Creal’s effectiveness relies heavily on Csmith seeds, while \tool is Csmith-free and only uses LLM-generated code through iterative synthesis. 
From this perspective, \tool can potentially extend to other languages where a mature Csmith-like generator is not available. 
In addition, as detailed below, \tool diverges from Creal in both the construction of its test database and the synthesis of complete programs, offering a wider applicability and greater flexibility.

\subsubsection{Enhancing Code Diversity via Code-Aligned Prompting}
For the database construction, the core contribution of \tool is the \textit{real-world code-aligned prompting} in Section~\ref{subsubsec:real-align}, which addresses the challenge of getting diverse and valid tests in LLM-based testing, which neither Creal nor other LLM-based compiler testing approaches can solve. 
For example, Creal is unable to transform the program shown in Figure~\ref{fig:alignment} (a), as it cannot handle non-numeric types in the \texttt{buf\_write} function syntactically, nor can it resolve the external function \texttt{realloc\_body} semantically. As a result, it achieves only a 5\% valid extraction rate. In contrast, \tool successfully processes such cases, achieving a valid rate of over 50\%.

\subsubsection{Generating Complex Programs through Iterative Synthesis}
For connecting program fragments, \tool presents a novel iterative synthesis methodology that connects different fragments through function call insertion and global variable sharing, as detailed in Section~\ref{subsec:online}. 
The function call insertion is indeed similar to Creal, but Creal’s design only supports \textbf{one-time} insertion with limited types, whereas \tool supports \textbf{unlimited insertion times} with richer type support. 
Other components, like global variable sharing and whole program construction, are \textbf{all unique} in \tool.
This limitation prevents Creal from covering more lines of code or detecting certain classes of bugs. We will demonstrate this in Section~\ref{sec:evaluation} through both coverage analysis and a case study highlighting \tool's ability to uncover complex bug patterns beyond Creal's reach.

\section{Implementation}\label{sec: impl}

This section details the implementation of real-world code-aligned prompting and explains how the synthesizer integrates into the overall fuzzing process.

\begin{figure*}[tp]
    \centering
    \includegraphics[width=0.97\linewidth]{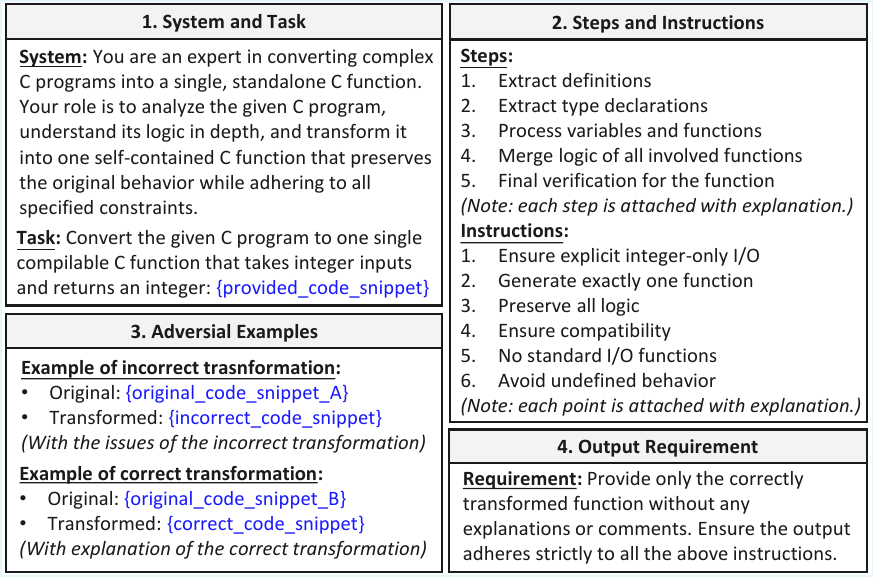}
    \vspace{-5pt}
    \caption{Pre-defined prompt guiding LLMs to transform real-world code.}
    \label{fig:prompt}
    \vspace{-10pt}
\end{figure*} 

\subsection{Prompt Design}\label{subsec:prompt}

Given a piece of code extracted from a real-world project, \tool uses a pre-defined prompt to guide the LLM to transform the code into a new version.
Figure~\ref{fig:prompt} shows the pre-defined prompt. It begins by defining the role of the LLM as an expert in writing programs and describing the overall task. The transformation process is guided by the \textit{chain-of-thought prompt} strategy~\cite{wei2022chainofthought}, which decomposes the task into ordered steps: understanding the code first, followed by generating the function, and finally verifying the result. 
To ensure real-world code alignment, detailed instructions are provided to enforce explicit numeric I/O, preserve logical structures, maintain compatibility, and avoid undefined behaviors. 
We also apply \textit{few-shot in-context learning}~\cite{brown2020fewshot}, offering examples of both correct and incorrect transformations as well as explanations.
The prompt concludes by specifying output format constraints to ensure that the transformed code can be extracted effectively from the response. 
By using this prompt with LLMs, we successfully transform \texttt{\{provided\_code\_snippet\}} into the formatted code. An example of this transformation has been shown in Figure~\ref{fig:alignment}. 
Note that other prompt designs may also work well. Our prompt here provides a useful example that is used in our implementation. Designing better prompting methods is interesting, but orthogonal to our work.

\subsection{Fuzzing Execution}\label{subsec:fuzzing}

Detecting crash bugs is straightforward.
If a compiler crashes when compiling a test program generated by \tool, we find a crash bug.
In order to find miscompilation bugs, 
like CSmith~\cite{yang2011csmith} and other tools derived from it~\cite{creal2024li}, \tool also employs randomized differential testing. 
We compute and print the checksum of the return values from the used functions and global variables to illustrate the behavior of the generated program. 
Specifically, we mutate a global variable by combining the return values of functions with the values of collected global variables. This mutation process is encapsulated within the main function, and a print statement is included to output the checksum of the modified global variable. 
For example, in Figure~\ref{fig:syn}, we will add more print statements in the main function to output the checksum of the global variable \texttt{g} and the return values of both \texttt{func1()} and \texttt{func2()}.
Finally, we compare the checksum values generated by different compilers and across various optimization levels to identify discrepant outputs. Discrepant outputs indicate the presence of a miscompilation bug in one of the compilers. We then reduce the generated program to a minimal example to manually decide which compiler is buggy and where to report the bug.

\section{Evaluation}\label{sec:evaluation}

In this section, we evaluate the effectiveness and design choices of \tool through the following research questions:

\begin{itemize}[labelwidth=!, labelindent=5pt, itemsep=3pt, topsep=2pt, leftmargin=15pt]
    \item \textbf{RQ1 (Bug finding and coverage analysis).} \emph{Is \tool effective in finding crash and miscompilation bugs in C compilers, and achieving high coverage?} 
    \item \textbf{RQ2 (Ablation analysis).} \emph{How important are the key components of \tool, including the code database, the choice of LLM, and the iteration number?}
    \item \textbf{RQ3 (Comparison with existing tools).} \emph{How does \tool compare with existing state-of-the-art LLM-based testing tools such as Fuzz4All, WhiteFox, and Creal?} 
    \item \textbf{RQ4 (Case study).} \emph{What types of bugs can \tool uncover, and how do they demonstrate the tool's unique strengths?}
\end{itemize}

\subsection{Experiment Setup} 
\noindent \textbf{Compiler Targets.} Our study primarily focuses on the two most widely used and mature C compilers, \ie, GCC (from \texttt{9366940} to \texttt{eb26b66}) and LLVM (from \texttt{b1560bd} to \texttt{029cb8a}). To ensure up-to-date evaluations, we update both compilers daily and utilize their latest versions for continuous testing. We apply five standard optimization levels, \ie, \texttt{-O0}, \texttt{-O1}, \texttt{-Os}, \texttt{-O2}, and \texttt{-O3}, across both compilers.

\vspace{2mm}
\noindent \textbf{Large Language Models.} We utilized ChatGPT-4o-mini, a fast and cost-effective lightweight model for LLM-based transformations. Specifically, we employ the \texttt{gpt-4o-mini-2024-07-18} checkpoint with a \texttt{max\_token} limit of 512 and a \texttt{temperature} setting of 0.7. 
Note that \tool is not limited to this specific model and also supports locally deployed LLMs. We will evaluate the potential impacts of different models in Section~\ref{subsec:rq2}.

\begin{figure}[tp]
\begin{minipage}{0.45\textwidth}
    \begin{table}[H]
    \centering
        \centering
        \footnotesize
        \caption{Status of Reported Bugs}
        \label{tab:status}
        \vspace{-10pt}
        \begin{tabular}{c|cc|c}
        \toprule
        \textbf{Status}       & \textbf{GCC} & \textbf{LLVM} & \textbf{Total} \\ \midrule
        Confirmed  & 0    & 2    & 2    \\
        Fixed      & 18   & 38   & 56   \\ 
        Duplicate  & 5    & 3    & 8   \\ \midrule
        Total     & 23   & 43   & 66   \\ \bottomrule
        \end{tabular}
    \end{table}

    \vspace{-10pt}
    
    \begin{table}[H]
        \centering
        \centering
        \footnotesize
        \caption{Symptoms of Reported Bugs}
        \label{tab:symptom}
        \vspace{-10pt}
        \begin{tabular}{c|cc|c}
        \toprule
        \textbf{Symptom}      & \textbf{GCC} & \textbf{LLVM} & \textbf{Total} \\ \midrule
        Crash & 7    & 29    & 36    \\
        Miscompilation      & 16   & 14   & 30  \\ \midrule
        Total     & 23   & 43   & 66   \\ \bottomrule
        \end{tabular}
    \end{table}
\end{minipage}
\hfill
\begin{minipage}{0.5\textwidth}
    \centering
    \includegraphics[width=0.97\linewidth]{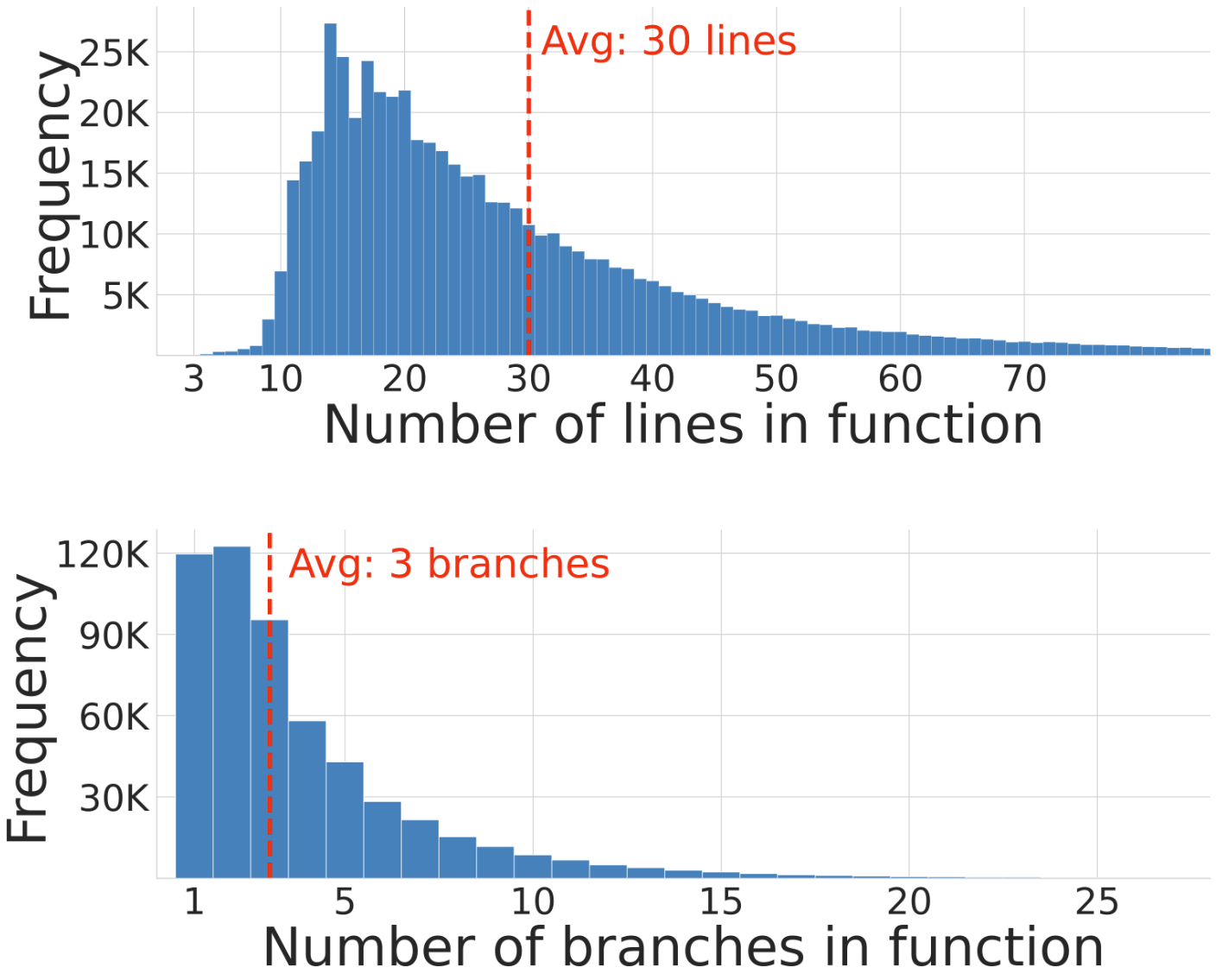}
    \vspace{-5pt}
    \caption{Distributions of functions in our database.
    }
    \label{fig:database}
\end{minipage}
\vspace{-5pt}
\end{figure}

\vspace{2mm}
\noindent \textbf{Code Database.} Our toolchain for database construction is partially built upon Creal~\cite{creal2024li}. To further enhance the function extraction capability from real-world projects, we extend its capabilities to support additional C features, such as structs, typedefs, and more. 
Instead of directly crawling open-source projects from GitHub, we selected AnghaBench~\cite{anghabench2021da}, which provides a vast collection of over 1,040,000 functions extracted from 146 open-source projects on GitHub. These projects include widely used software such as Linux, Redis, and Nginx, among others. 
After the offline code collection described in Section~\ref{subsec:offline}, we constructed a database with 553,246 functions, around 53\% of all functions in AnghaBench. 
Most discarded functions are due to the LLM not extracting an invalid function from them. 
Although the LLM does not always follow our instructions, our filtering process ensures that only valid code is preserved. 
Among these discarded functions, the most common reason is the violation of the first instruction in Figure~\ref{fig:prompt}, with an invalid I/O function rate of 14.2\%. 
The whole database construction costs us \$394 in invoking ChatGPT-4o-mini.
Figure~\ref{fig:database} shows the distribution of lines of code and branches in our database. The majority of functions contain fewer than 60 lines of code and fewer than 10 branches. On average, each function consists of approximately 30 lines of code and 3 branches. 
As has been discussed in Section~\ref{subsubsec:database}, these functions alone can not find any compiler bugs.

\vspace{2mm}
\noindent \textbf{Environment.} We conducted all our evaluations on one Linux server running Ubuntu 20.04 LTS. It is equipped with an AMD EPYC 7742 64-core CPU and 256GB RAM.

\vspace{2mm}
\noindent \textbf{Testing Process.} We conducted the fuzzing process continuously on a dedicated server to ensure thorough and uninterrupted testing. In each round, \tool selects a single function as the seed and generates 10 mutant versions of it.
For each synthesized test case, we compile and execute the program using both GCC and LLVM, leveraging their outputs as the test oracle, as discussed in Section~\ref{subsec:fuzzing}. 
Each fuzzing process is constrained to 1 GB of memory and a 200-second compilation timeout to prevent resource exhaustion and ensure fairness. 
If a miscompilation or runtime crash is detected, we employ C-Reduce~\cite{online:creduce} to minimize the faulty program, isolating the root cause of the issue. Finally, we submit a detailed bug report, including the reduced test case, to the respective compiler developers for further diagnosis.

\subsection{RQ1: Bug Finding and Coverage Analysis} \label{subsec: rq1}
Bug-finding capability and program coverage are fundamental metrics for evaluating the effectiveness of a testing approach. We first present the bug-finding results of \tool, followed by an analysis of the coverage achieved by its synthesized programs. 

\vspace{2mm}
\noindent \textbf{Baseline: individual function testing.} 
We conducted an experiment using functions directly from our database to test the latest versions of GCC and LLVM. The results were conclusive --- \emph{no compiler bugs were detected}. 
This outcome is not surprising. While our database collectively contains diverse program features, each individual function implements only a limited subset of these features. 
Modern compiler optimizations operate on complex interactions between program elements, often requiring specific combinations of features to trigger bugs. 
Programs that expose compiler bugs typically contain intricate control flows, data dependencies, and feature interactions that simple, isolated functions lack. 

\vspace{2mm} 
\noindent \textbf{Number of Bugs.} Table~\ref{tab:status} provides the status summary of all reported bugs identified by \tool. In total, \tool has reported 66 bugs; 58 of them (88\%) have been confirmed as previously unknown and new bugs, and 56 bugs (85\%) have already been fixed.
Specifically, GCC developers have fixed 100\% (18/18) of the reported bugs, while LLVM developers have fixed 95.0\% (38/40) of them. 
This highlights the effectiveness of \tool in identifying critical issues and the willingness of both GCC and LLVM developers to address our reported bugs. 
Since compiler maintainers, users, and testers are also testing compilers, \tool reported 8 duplicate bugs that were concurrently identified by them. 
\emph{Nevertheless, the significant number of new bugs identified by \tool demonstrates its strong bug-finding capability.}

\vspace{2mm}
\noindent \textbf{Symptoms of Bugs.} Table~\ref{tab:symptom} summarizes the symptoms of our reported bugs. These bugs can be categorized into two main types: 
(1) \textit{Crash}: This type of bug occurs when the compiler encounters an internal error during the compilation process, typically due to assertion failures or runtime failures. 
(2) \textit{Miscompilation}: In this case, the compiler implicitly generates incorrect code without any observable consequences. Such bugs are hard to detect and are the most concerning type of bugs in compilers~\cite{chen2021compilersurvey}.
As shown in the table, \emph{nearly half of the bugs are miscompilation bugs}, demonstrating the strong capability of \tool in detecting hard-to-detect bugs.
Detecting such miscompilation bugs requires valid and feature-rich testing programs.
As a comparison, none of the existing LLM-based compiler testing tools~\cite{yang2024whitefox, xia2024fuzz4all} can find any miscompilation bugs in either GCC or LLVM.

\begin{figure}[tp]
    \centering
    \begin{minipage}{0.48\textwidth} 
        \centering
        \includegraphics[width=\linewidth]{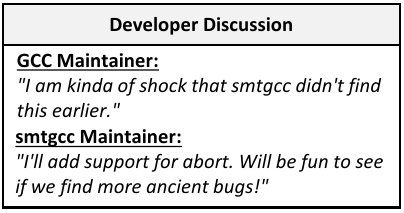}
        \caption{Developers' discussions on bug report \url{https://gcc.gnu.org/bugzilla/show_bug.cgi?id=118915}.}
        \label{fig:conversation}

        \vspace{5pt}

        \centering
        \includegraphics[width=\linewidth]{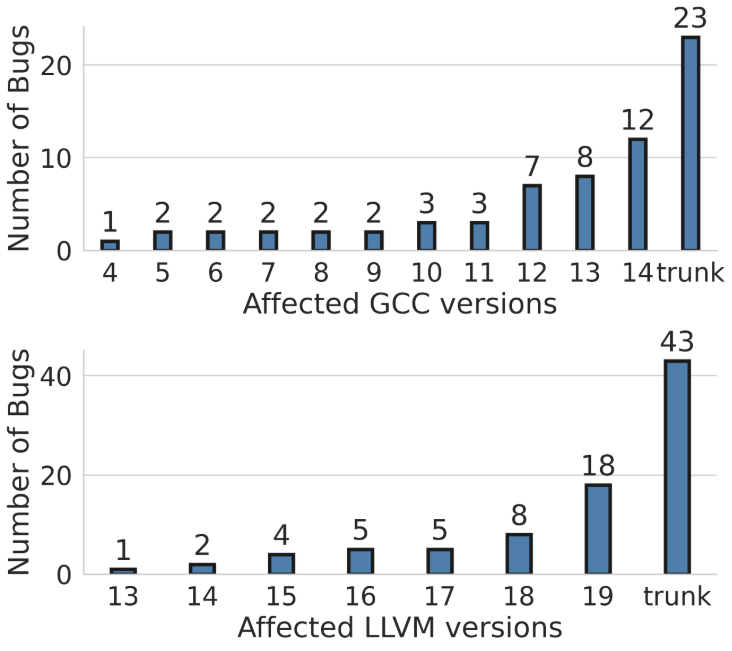}
        \caption{Stable compiler versions affected by our reported bugs.}
        \label{fig:affected-versions}
    \end{minipage}
    \hfill
    \begin{minipage}{0.48\textwidth}
        \centering
        \footnotesize
        \renewcommand{\arraystretch}{1.2}
        \setlength{\tabcolsep}{2pt}
        \captionof{table}{Affected LLVM components.}
        \label{tab:affected-llvm-components}
        \vspace{-5pt}
        \begin{tabular}{lr}
            \toprule
            \textbf{Component} & \textbf{\#Bugs} \\ 
            \midrule
            Loop Transformations & 15 \\
            Vectorization Optimization & 8 \\
            Peephole Optimizations & 4 \\
            SLP Vectorization & 4 \\
            Backend & 2 \\
            Selection DAG & 2 \\
            Scalar Evolution Analysis & 2 \\
            Dead Store Elimination & 1 \\
            Dominance based Optimizations & 1 \\
            Induction Variable Transformations & 1 \\
            Loop Invariant Code Motion & 1 \\
            \bottomrule
        \end{tabular}

        \vspace{10pt} 

        \footnotesize
        \renewcommand{\arraystretch}{1.2}
        \setlength{\tabcolsep}{2pt}
        \captionof{table}{Affected GCC components.}
        \label{tab:affected-gcc-components}
        \vspace{-5pt}
        \begin{tabular}{lr}
            \toprule
            \textbf{Component} & \textbf{\#Bugs} \\ 
            \midrule
            Peephole Optimizations & 8 \\
            Backend & 3 \\
            CFG Transformations & 2 \\
            Constant Propagation & 2 \\
            Loop Invariant Motion & 2 \\ 
            IPA constant propagation & 1 \\
            Number of Iterations Analysis & 1 \\
            Predictive Commoning & 1 \\
            Value Numbering & 1 \\
            Value Range Analysis & 1 \\
            Vectorization & 1 \\
            \bottomrule
        \end{tabular}
    \end{minipage}
\end{figure}

\vspace{2mm}
\noindent \textbf{Importance of bugs.} 
To assess the impact of the discovered bugs, we tested each bug-triggering program to see if they can trigger crashes or miscompilations on earlier stable compiler versions.
Figure~\ref{fig:affected-versions} shows the number of bugs affecting different compiler versions.
The result indicates that \tool is highly effective at uncovering \emph{long-latent} bugs --- issues that have persisted undetected for years despite extensive compiler testing efforts. 
Notably, we identified 7 bugs that affected GCC versions predating GCC-12 and 8 bugs that impacted LLVM versions released before LLVM-18. Given that these compiler versions have been released from 1 to 10 years ago, the longevity of these undetected bugs underscores their criticality.
Remarkably, one particular GCC miscompilation bug was traced back to a code change that has existed since at least 2006. 
This demonstrates that the bug remained unnoticed for nearly two decades, highlighting the limitations of existing testing methodologies. 
Since there are numerous compiler testing efforts in both academia and industry,
\emph{the fact that these long-latent bugs have evaded all previous testing techniques further highlights the exceptional bug-finding capability of \tool}.

Beyond exposing long-hidden issues, some of our discovered bugs also provide valuable insights for the improvement of future compilers or related tools.
For example, one miscompilation bug\footnote{\url{https://gcc.gnu.org/bugzilla/show_bug.cgi?id=118638}} was found to have implications for future, yet-to-be-developed versions of GCC.
Furthermore, discussions regarding the root cause of a long-latent GCC miscompilation bug led to unexpected reactions from developers. Figure~\ref{fig:conversation} shows the comments we received for one of our reported bugs. A GCC maintainer expressed the surprise that our reported bug had not been detected by smtgcc~\cite{online:smtgcc}, which is designed to guarantee the detection of the reported bug. In response, the smtgcc maintainer acknowledged the limitation and stated that he would introduce additional support to detect similar \emph{ancient} bugs in future analyses. 

\vspace{2mm}
\noindent \textbf{Affected compiler components.} 
Table~\ref{tab:affected-gcc-components} and Table~\ref{tab:affected-llvm-components} summarize the compiler components affected by the bugs we identified in GCC and LLVM, respectively.
These components were determined based on the diagnostic information and fix messages provided by compiler maintainers.
As shown in the results, \tool is capable of uncovering a diverse range of bugs. In both GCC and LLVM, many of the bugs are related to loop transformations and peephole optimizations, which is consistent with findings from prior empirical studies~\cite{zhou2021optimization}.
To offer a more comprehensive understanding, we will present representative bug cases in Section~\ref{sec:case-study}, demonstrating the distinctive characteristics of our synthesized programs and their effectiveness in compiler testing.

\setlength{\tabcolsep}{2pt}
\begin{table}[]
\centering
\footnotesize
\begin{minipage}{0.67\linewidth}
    \centering
    \renewcommand{\arraystretch}{1.7}
    \begin{tabular}{lllll}
    \toprule
    \textbf{Compiler}              & \textbf{Generator}            & \textbf{FC}     & \textbf{LC}       & \textbf{BC}       \\ \midrule
    \multirow{3}{*}{\textbf{GCC}}  
        & Seeds (1,000)                  & 31.9\%          & 27.0\%            & 16.0\%            \\
        & Functions (10,000)            & 36.0\% (+3,575) & 33.3\% (+55,916)  & 20.7\% (+51,419)  \\
        & \textbf{\tool (10,000)}       & \textbf{38.9\% (+6,103)} & \textbf{39.5\% (+110,945)} & \textbf{25.9\% (+108,308)} \\ 
    \midrule
    \multirow{3}{*}{\textbf{LLVM}} 
        & Seeds (1,000)                  & 24.3\%          & 31.7\%            & 16.0\%            \\
        & Functions (10,000)            & 26.4\% (+1,979) & 34.6\% (+47,522)  & 20.3\% (+30,008)  \\
        & \textbf{\tool (10,000)}       & \textbf{27.6\% (+3,110)} & \textbf{36.6\% (+80,295)}  & \textbf{23.8\% (+54,433)}  \\ 
    \bottomrule
    \end{tabular}
    \vspace{5pt}
    \caption{Line coverage (LC), function coverage (FC), and branch coverage (BC) of GCC and LLVM.}
    \label{tab:cov}
\end{minipage}
\hfill
\begin{minipage}{0.28\linewidth}
    \centering
    \includegraphics[width=\linewidth]{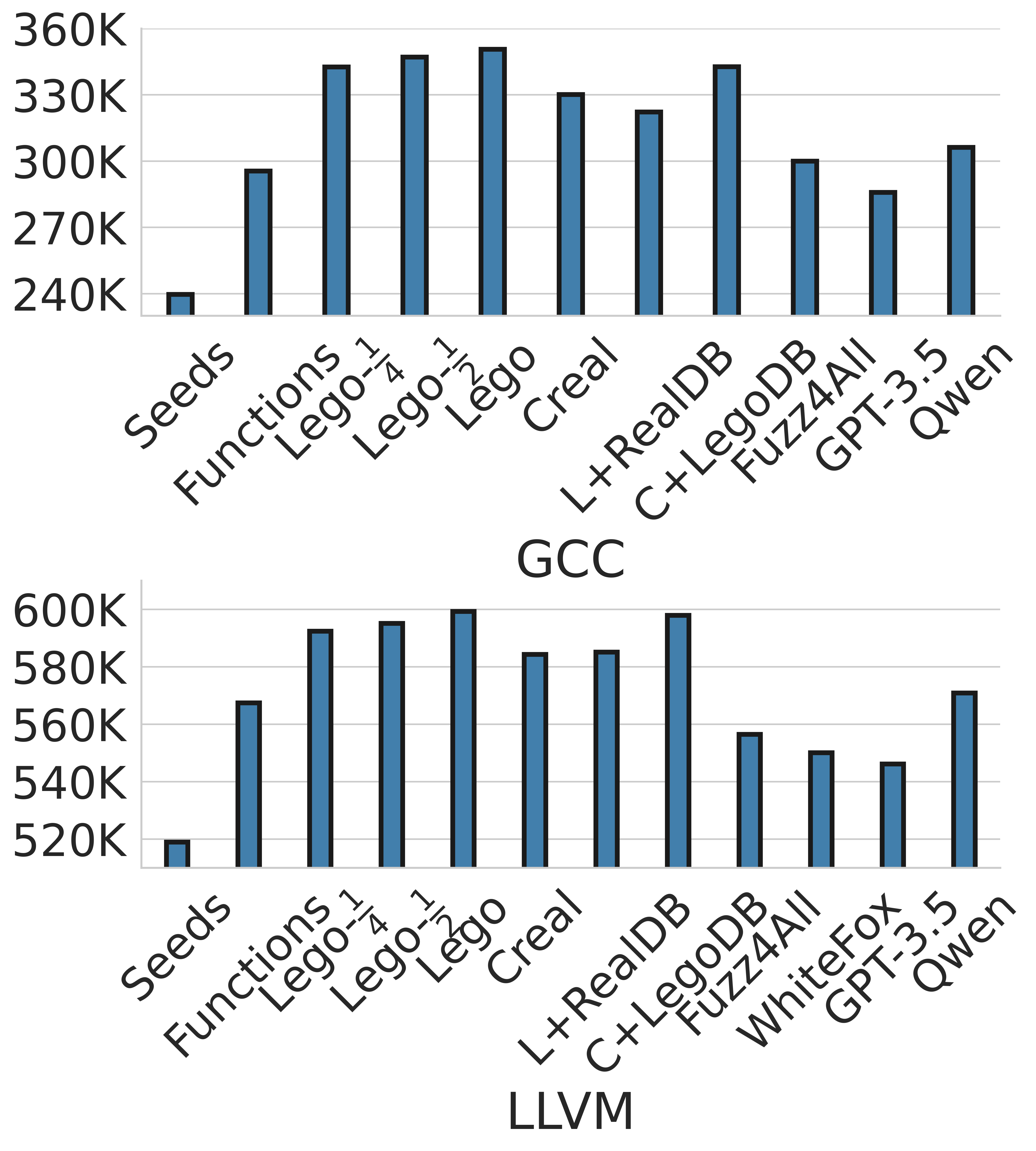}
    \vspace{-10pt}
    \captionof{figure}{Line coverage of different variants}
    \label{fig:coverage}
\end{minipage}
\vspace{-10pt}
\end{table}


\vspace{2mm}
\noindent \textbf{Coverage and generation speed.} 
We perform the coverage analysis and track the generation time throughout the fuzzing process to assess whether \tool can effectively explore more parts of the compilers and how efficient it is.
Starting from 1,000 randomly selected seed functions, we let \tool generate 10 programs for each seed, resulting in 10,000 programs produced by \tool.
We then measure function coverage, line coverage, and branch coverage of these programs across GCC and LLVM. 

Table~\ref{tab:cov} and Figure~\ref{fig:coverage} present the results of our coverage analysis. 
The ``Seed'' rows in Table~\ref{tab:cov} and bars in Figure~\ref{fig:coverage} show the coverage of the 1,000 seed functions that \tool starts from.
The ``Functions'' rows and bars show the coverage of 10,000 randomly selected functions from our database. That is, we directly use these functions individually as testing programs. 
The ``\tool'' rows and bars show the coverage of the 10,000 programs generated by \tool.
It is not surprising that \tool significantly enhances coverage compared to the seed functions. Specifically, \tool increases line coverage in GCC by 12.5\%, corresponding to 110,945 additional lines, and in LLVM by 2.9\%, covering 80,295 more lines. 
Similar trends are observed for function coverage and branch coverage.
Compared to ``Functions'', \tool achieves substantially higher coverage, indicating that iterative synthesis plays a crucial role in generating more complex programs that engage a wider range of compiler features. 

We also logged the time used to generate the 10,000 programs by \tool. In total, \tool generated these 10,000 programs in 193 seconds, averaging 0.02 seconds per program. 
This high efficiency indicates that program generation is not a bottleneck for \tool. In fact, we observed that most of the time during compiler testing is spent on compiling the generated programs, which can often take several seconds per program.

\subsection{RQ2: Ablation Analysis}\label{subsec:rq2}
We study the impact of the code database, the choice of LLM, and the iteration number on the effectiveness of \tool, using coverage as the primary metric. 

\vspace{2mm}
\noindent \textbf{Importance of code database.} 
The effectiveness of \tool is closely related to the quality of the code database. A consequent question is \emph{whether the quality and size of the code database contribute to the effectiveness of \tool}.
To answer this question, we conduct a coverage analysis on different variants of the database:

\begin{itemize}[itemsep=3pt] 
    \item \emph{\tool-$\frac{1}{2}$}: We reduce the size of the code database used in \tool by half. Specifically, we randomly select half of the available functions and use this smaller subset during generation of 10,000 cases. 
    \item \emph{\tool-$\frac{1}{4}$}: We further halve the database, reducing it to a quarter of its original size to generate 10,000 cases.
    \item \emph{\tool-Creal}: We apply Creal’s original database to guide \tool in generating 10,000 test cases. The resulting coverage is shown in the ``L+RealDB'' bar in Figure~\ref{fig:coverage}. 
    \item \emph{Creal-\tool}: We use \tool's code database to generate 10,000 test cases under the Creal framework. The ``C+LegoDB'' bar in Figure~\ref{fig:coverage} shows the resulting coverage. 
\end{itemize}

The third to fifth bars in Figure~\ref{fig:coverage} show the results of using different size of \tool's code database. We can observe a clear trend: as the code database size increases, so does the achieved line coverage.
\tool outperforms \tool-$\frac{1}{2}$ and \tool-$\frac{1}{2}$ outperforms \tool-$\frac{1}{4}$ in both GCC and LLVM, indicating that a larger code database provides more diverse features. 
This also suggests that future work on establishing a larger code database can potentially further improve \tool.
We observe that applying Creal's database to guide \tool (``L+RealDB'') leads to reduced coverage compared to \tool with its own database, highlighting the importance of our LLM-driven offline phase and real-world code-aligned prompting strategy.
Meanwhile, when Creal is guided by our database (``C+LegoDB''), it achieves substantially higher coverage than using its original database (``Creal''), as shown in Figure~\ref{fig:coverage}.
This demonstrates that our code database is also broadly effective across different frameworks.



\begin{figure*}[tp]
    \centering
    \begin{minipage}{0.45\linewidth}
        \centering
        \includegraphics[width=\linewidth]{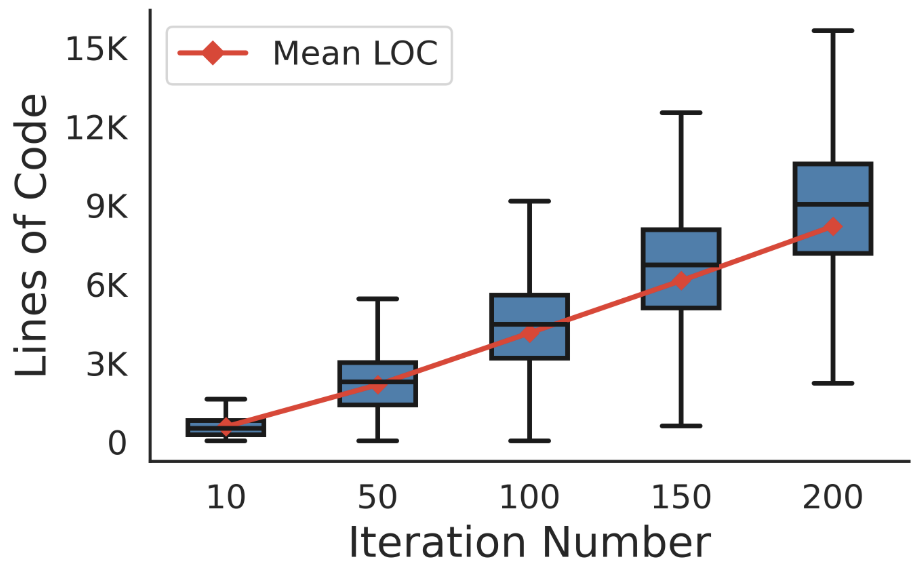}
        \vspace{-15pt}
        \caption{Relationship between iteration number and lines of code.
        }
        \label{fig:trend}
    \end{minipage}
    \hfill
    \begin{minipage}{0.45\linewidth}
        \centering
        \includegraphics[width=\linewidth]{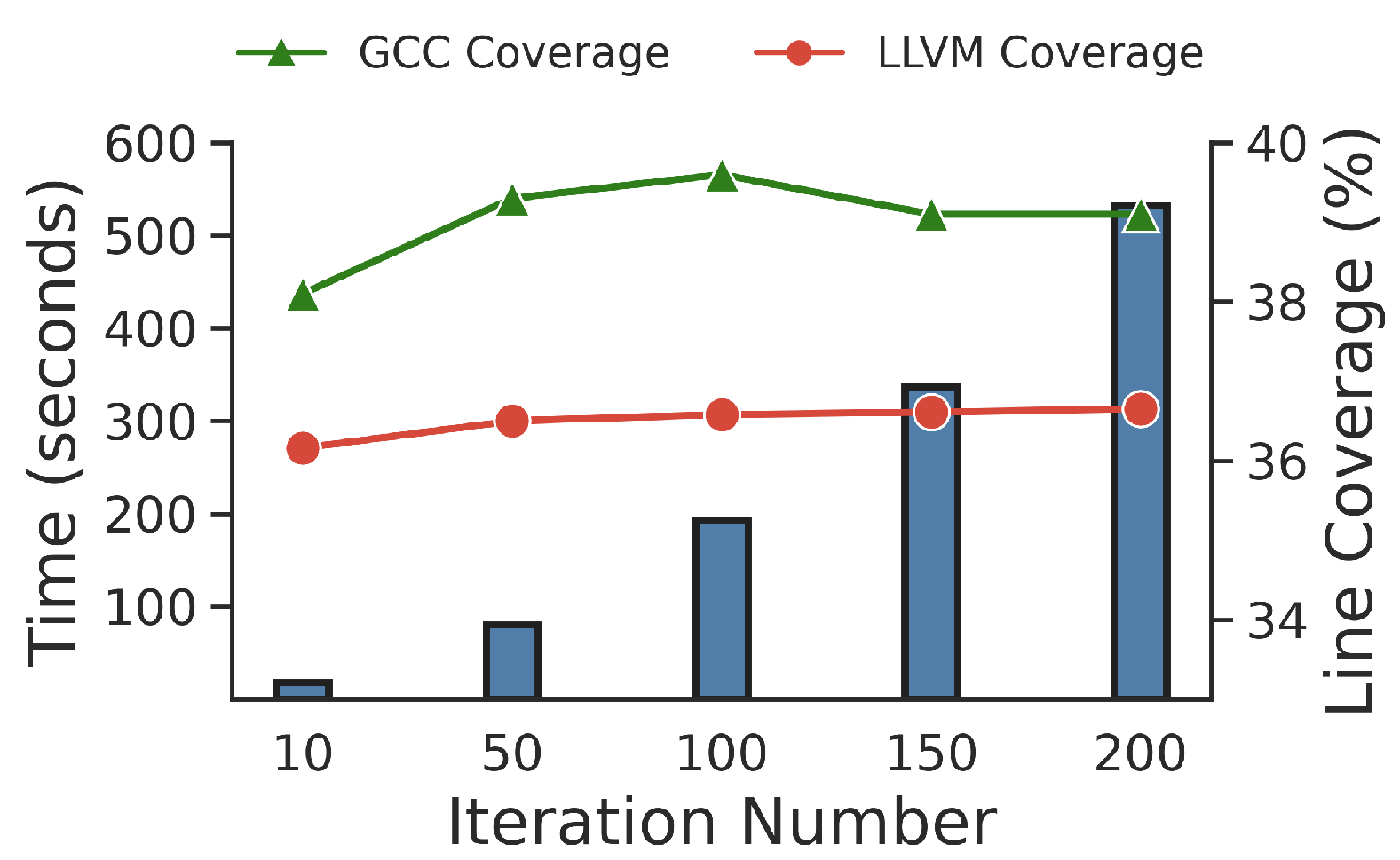}
        \vspace{-15pt}
        \caption{Relationship between iteration number, generation time, and line coverage.}
        \label{fig:iteration}
    \end{minipage}
    \vspace{-10pt}
    \label{fig:comparison}
\end{figure*}


\vspace{2mm}
\noindent \textbf{Alternative Large Language Models.} 
In our current implementation, we utilize ChatGPT-4o-mini as the underlying LLM. 
Although the choice of LLM is orthogonal to the design of our framework, we aim to explore the adaptability and generalizability of our framework across different models. 
We select ChatGPT-3.5-turbo~\cite{online:3.5-turbo} and Qwen2.5 Coder 32B Instruct~\cite{online:qwen} for evaluation. 
Since GPT-3.5-turbo is a legacy model, we aim to evaluate whether our framework can effectively adapt to a less powerful model. 
Since Qwen Coder is a well-known code-specific LLM, we aim to evaluate whether a specialized model can generate a more expressive code database. 

The ``Functions'' rows in Table~\ref{tab:cov} show the coverage of the 10,000 functions in the database used by \tool. These functions are generated by GPT-4o-mini by transforming 10,000 real-world code snippets. 
With the same pipeline, we ask each of the two new LLMs to generate 10,000 functions. Finally, we evaluate the coverage of these two sets of new functions.

The ``GPT-3.5'' and ``Qwen'' bars in Figure~\ref{fig:coverage} show the coverage of the 10,000 functions generated by GPT-3.5-turbo and Qwen Coder, respectively.
As expected, GPT-3.5-turbo achieves lower coverage compared to the more advanced GPT-4o-mini (``Functions''). 
Qwen Coder, on the other hand, achieves higher coverage compared to GPT-4o-mini, indicating that a specialized model can generate a more expressive code database and potentially improve the performance of \tool.
Nevertheless, our design of \tool is loosely coupled with the choice of LLM, and the prompt can be easily adapted to better suit specific LLMs. Our core contribution is the design of a novel testing framework, which is orthogonal to the choice of LLM.


\vspace{2mm}
\noindent \textbf{Impact of iteration number.} 
In Algorithm~\ref{alg:generate}, \tool uses the iteration number $\mathcal{N}$ to control the number of synthesis iterations during the online phase. 
To understand the impact of $\mathcal{N}$, we set $\mathcal{N}$ to 10, 50, 100, 150, and 200.
For each $\mathcal{N}$, we generate 10,000 programs.
We then compute the average lines of code (LOC) across the programs, the time of generating 10,000 programs, and the line coverage of these programs. 

Figure~\ref{fig:trend} shows the trend in LOC as $\mathcal{N}$ increases. There is a clear upward trend in LOC, with the maximum LOC exceeding 15,000 when $\mathcal{N}$ reaches 200. This is expected as a higher $\mathcal{N}$ leads to more functions being selected for synthesis.
Figure~\ref{fig:iteration} shows that more iterations require more time to generate programs, but the coverage does not increase much after $\mathcal{N}$ reaches 100.
The main reason is that additional functions introduced by more iterations do not contribute meaningfully to the overall coverage, as most of the relevant code paths have already been explored. 
They may even hinder compiler optimization, as evidenced by the noticeable decline in GCC coverage.
Therefore, for practical real-world fuzzing applications, setting 
$\mathcal{N}$ to 100 offers a good balance between a reasonable generation time and a high coverage.

\subsection{RQ3: Comparison with Existing Tools}\label{subsec:comparison}


We compare \tool with Fuzz4All~\cite{xia2024fuzz4all}, WhiteFox~\cite{yang2024whitefox}, and Creal~\cite{creal2024li}, three state-of-the-art compiler testing frameworks. 
While Fuzz4All and WhiteFox represent recent LLM-based approaches, Creal integrates real-world code with traditional fuzzing via Csmith. 

\para{Bug-finding analysis.} 
We manually collected and analyzed all the reported GCC and LLVM bugs by Fuzz4All and WhiteFox.
The results show that \emph{neither Fuzz4All nor WhiteFox has discovered any miscompilation bugs.} 
In particular, Fuzz4All focused on \texttt{g++}, GCC's C++ compiler. In total, Fuzz4All discovered only eight confirmed bugs with half of them being fixed. Notably, all of them are related to \texttt{g++} frontend issues, rather than compiler optimizations. 
WhiteFox, on the other hand, has identified two bugs in LLVM --- one is a backend-related crash, and another is a crash in error diagnostics. 
These findings suggest that both tools have very limited effectiveness in testing C compilers, as their discovered bugs do not touch the core compiler optimizations.
Since Creal has extensively tested the GCC and LLVM, the fact that \tool discovered many long-latent bugs demonstrates the strong complementary bug-finding capability of Creal. We will discuss such cases in Section~\ref{sec:case-study} to show the distinctive characteristics of discovered bugs.

\para{Coverage analysis.} 
We generate 10,000 test cases using Fuzz4All, WhiteFox and Creal for evaluation. Since WhiteFox is designed to generate C++ code, which is then converted into LLVM IR instead of directly targeting C, we only measure its LLVM coverage. 

For LLM-based tools, as shown in Figure~\ref{fig:coverage}, Fuzz4All achieves significantly lower line coverage in GCC compared to \tool, with a coverage gap of over 50,000 lines of code. In LLVM, both tools exhibit even lower coverage than the raw function-level inputs used by \tool prior to the synthesis process, highlighting their limited effectiveness in covering compiler optimizations.  
For Creal (``Creal''), \tool still outperforms it substantially, demonstrating the advantage of our proposed unique methods. We further observe that using the same code database (``L+RealDB''), \tool still achieves higher coverage than Creal, highlighting the effectiveness of its iterative synthesis strategy.

\para{Generation speed analysis.}
The time cost of generating 10,000 test programs for Fuzz4All and WhiteFox is 12,121 and 28,284 seconds, respectively.
Compared to \tool's default configuration with $\mathcal{N} = 100$ (193 seconds), Fuzz4All and WhiteFox are considerably slower, taking 62 times and 146 times longer, respectively.
Although all tools rely on LLMs, the offline and online decoupling design of \tool allows it to generate test programs at a much faster speed.
We also include Creal in this comparison, which takes 13,997 seconds to generate the same number of programs. 
\tool eliminates the dependency on Csmith-generated seeds by directly composing functions from code database, which enables not only faster generation but also greater portability to languages without mature seed generators.

\begin{figure*}[tp]
    \centering
    \includegraphics[width=\linewidth]{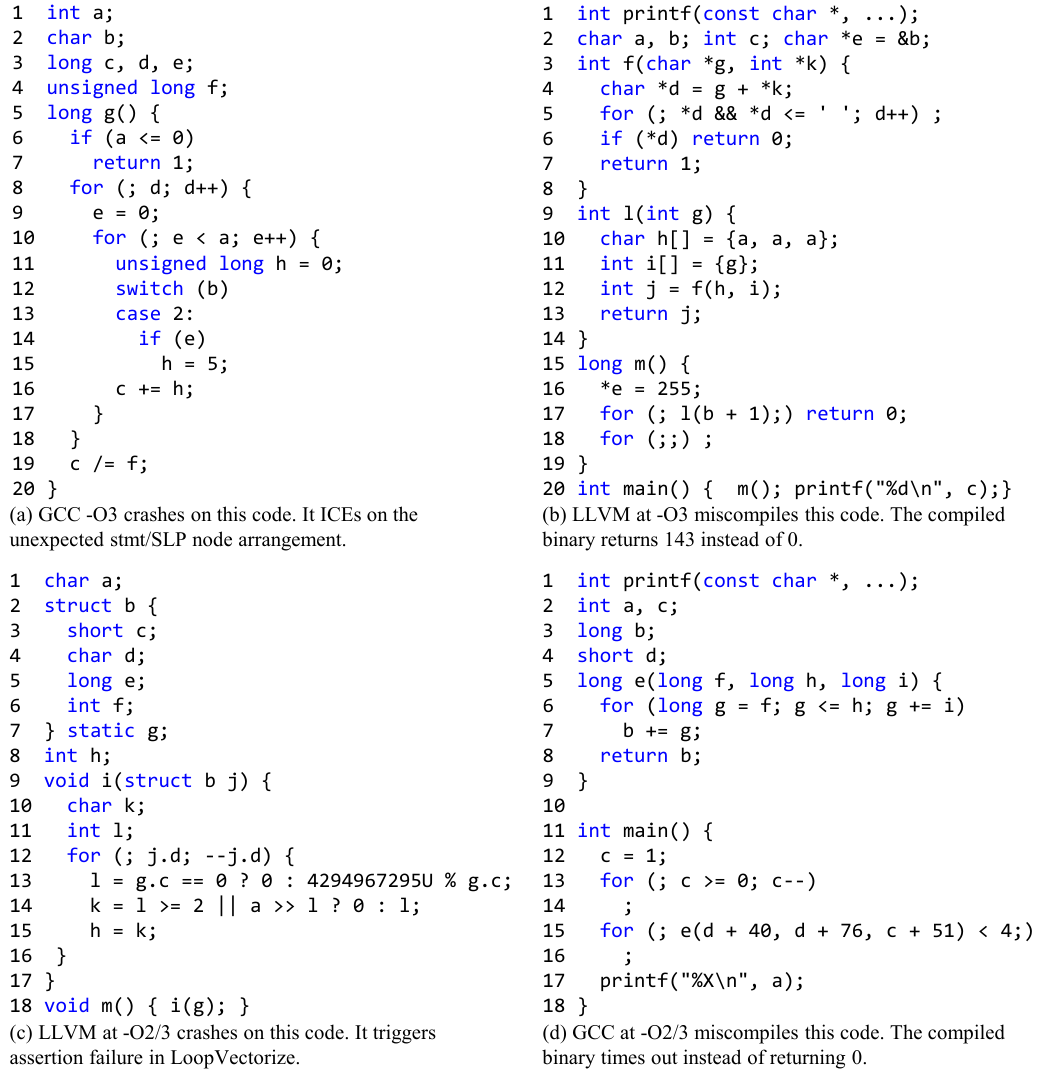}
    \vspace{-10pt}
    \caption{Sample reduced programs that trigger compiler bugs.}
    \label{fig:case-stduy}
\end{figure*} 

\subsection{RQ4: Case Study} \label{sec:case-study}

This section presents four cases to demonstrate that \tool can discover deep bugs in compilers. Note that all programs are reduced and simplified from the original ones for better readability.

\smallskip\noindent\textbf{Figure~\ref{fig:case-stduy} (a):} 
The original bug-triggering program generated by \tool contains 4,945 lines and 62 functions.
This reduced program triggers a crash in GCC with the -O3 optimization flag. The issue occurs because GCC mistakenly classifies a reduction pattern in the outer loop when, in fact, the variable is not used outside the loop. Specifically, in line 16, the variable \verb|c| is incremented within the inner loop, which GCC misinterprets as a reduction operation that spans both loops. 
This leads to an assertion failure during the vectorization pass, resulting in a compiler crash. 

\smallskip\noindent\textbf{Figure~\ref{fig:case-stduy} (b):} 
The original bug-triggering program generated by \tool contains 6,697 lines and 91 functions.
This reduced program triggers a miscompilation in LLVM due to an incorrect transformation performed by the InstCombine pass. The issue arises from an invalid optimization applied immediately after inlining the function \verb|l()| into \verb|m()|.
Specifically, in line 5, the function \verb|f()| computes a pointer offset using \verb|g + *k|, where \verb|g| is an array and \verb|k| is an integer pointer. However, after inlining, LLVM incorrectly simplifies the \verb|sext| operation, leading to an incorrect \verb|getelementptr| (GEP) offset calculation in line 10. This results in a miscomputed pointer access, which causes undefined behavior when dereferencing \verb|d| in line 6. 
This miscompilation causes the compiled binary to return 143 instead of 0.
This program features a long and semantically meaningful function call chain starting from \texttt{m()}, which is rarely observed in Creal's bug cases.
Such patterns emerge naturally from \tool's unique iterative program synthesis, where complex interactions between function calls and global variables are progressively constructed. 

\smallskip\noindent\textbf{Figure~\ref{fig:case-stduy} (c):} 
The original bug-triggering program generated by \tool contains 3,989 lines and 107 functions.
This reduced program triggers a crash in LLVM.
The bug occurs due to a discrepancy between the new VPlan cost model and the legacy cost model. Specifically, in line 12, the loop iterates when \verb|j.d| is nonzero, decrementing \verb|j.d| in each iteration. Inside the loop, in line 13, the expression \verb|4294967295U % g.c| is computed, which results in an undefined behavior if \verb|g.c| is zero.  
LLVM's loop vectorization planner attempts to determine the best vectorization factor, but during cost analysis, an inconsistency arises between the two cost models. This results in an assertion failure in \verb|LoopVectorize| due to a mismatch in the expected vectorization behavior. 

\smallskip\noindent\textbf{Figure~\ref{fig:case-stduy} (d)} 
The original bug-triggering program generated by \tool contains 6,485 lines and 24 functions.  
This reduced program triggers a miscompilation bug in GCC. The root cause is in the loop iteration analysis. Specifically, in line 7, the function \verb|e()| accumulates values into the global variable \verb|b| using a loop in steps of \verb|i|. However, in line 15, the second loop in \verb|main()| calls \verb|e(d + 40, d + 76, c + 51)|, with \verb|c| being decremented to \verb|-1| in the previous loop. Since \verb|c| influences the step size (\verb|i| in \verb|e()|), the compiler performs an invalid transformation of the loop bound, leading to an incorrect evaluation of the loop exit condition.
GCC incorrectly optimizes the loop termination check, replacing a less-than comparison (\verb|g <= h|) with a non-equal comparison (\verb|g != h|), which leads to an unintended infinite loop depending on the initial values.

\section{Discussion}\label{sec:discussion}

\para{\tmark~Alternative prompting methods.}
\tool adopts a flexible design that allows for prompt customization, enabling it to better align with specific models. As Section~\ref{subsec:rq2} indicates, the current prompt is particularly effective for models that excel at processing natural language instructions. 
However, it may not be as suitable for task-specific models, such as those specialized in code generation. By adjusting the prompt, we can optimize the transformation process to accommodate different types of language models and enhance their performance in generating diverse code structures.
Better prompt design and better LLM models can improve the quality of the constructed code database, further enhancing the effectiveness of \tool.

\para{\tmark~More function signatures.}
Our current approach leverages LLMs to generate numeric functions only.
The core reason is to simplify the iterative synthesis process. Using numeric values for both function inputs and outputs and global variables eases our engineering effort in connecting different codes together. Supporting more types is theoretically doable, but it would require additional engineering efforts. Furthermore, even for numeric functions, the function bodies contain a much divergent range of types, such as strings and user-defined structs, and thus, the overall expressiveness of the generated programs is not affected.

\para{\tmark~Alternative sources of code snippets.}
\tool shows, for the first time, that LLMs can be used to find deep compiler bugs. 
Given a constructed code database, the online synthesis component of \tool can generate testing programs. It is theoretically possible to use methods other than LLMs to construct the code database. For example, we can use historical bug-triggering programs~\cite{skeleton2017pldi} or real-world functions directly~\cite{creal2024li}. However, we argue that these sources are not as extensive as LLMs. 
Compared to real-world programs, historical bug-triggering programs are far less diverse. Directly using real-world programs is also constrained by their complexity and uncertainty. For example, Creal~\cite{creal2024li} only manages to collect fifty thousand functions from one million real-world programs.
In comparison, we can generate over half a million functions.
Therefore, coupled with LLMs, \tool can be more effective in discovering deep bugs.

\para{\tmark~Beyond function level synthesis.}
\tool chains multiple functions together to form a testing program. One may wonder whether working at the granularity of function would lead to a loss of diversity in the generated programs.
We argue that these programs are expressive enough to cover a wide range of compiler behaviors:
(1) Our bug-finding results in Section~\ref{sec:evaluation} have already shown that \tool can discover deep bugs in various compiler optimizations, and (2) most compiler optimizations work at the function level~\cite{llvmpass, gccpass}, ensuring that our testing programs can exercise a wide range of compiler optimizations.
Extending \tool to support other levels of program synthesis is an interesting direction for future work and does not affect the core contribution of \tool.
For example, when building dependencies between two functions, instead of generating a function call, we can directly merge one function body into another to form a larger function.
However, this is similar to the existing practice of inlining functions during compilation. Adding ``\_\_attribute\_\_((always\_inline))'' directive to function definitions can also achieve a similar effect.

\para{\tmark~Beyond C compiler testing.}
\tool describes a new paradigm of compiler testing, \ie, using LLMs to generate building blocks first and thus use them to synthesize testing programs.
This paper provides the first proof-of-concept implementation of this new paradigm on testing C compilers.
Implementing \tool takes much less engineering effort than writing a program generator from scratch.
This opens up many exciting opportunities for testing compilers for various programming languages, such as Rust, where a reliable and efficient program generator is not yet available.

\section{Related Work}
\subsection{Generation-based compiler testing.} 
Significant efforts have been dedicated to developing automated program generators for compiler testing.
Csmith~\cite{yang2011csmith} is one of the most widely used program generators for detecting C/C++ compiler bugs. It can generate a large number of test programs covering a broad subset of the C language while avoiding undefined behaviors, making it an essential tool for compiler validation. 
CLsmith~\cite{christopher2015opencl} is inspired by Csmith, which is a program generator for OpenCL compilers and has six modes for generation. 
Morrisset \etal~\cite{morisset2013concurrency} extended Csmith by incorporating support for mutexes, atomic variables, and system calls for locking and unlocking mutexes, enabling the detection of C/C++ concurrency bugs.
YARPGen~\cite{livinskii2020yarpgen} and its successor, YARPGen v2~\cite{livinskii2023yarpgen}, are modern program generators specifically designed to test scalar and loop optimization bugs.

While these tools have been effective in discovering numerous compiler bugs, they are rule-driven and may eventually reach a saturation point~\cite{amalfitano2015saturation}, where they struggle to uncover new bugs in a given compiler. This limitation arises from the predefined constraints embedded in these generators, which restrict their ability to explore certain aspects of compiler behavior.
\tool addresses this limitation by leveraging real-world programs instead of relying solely on predefined generation rules. By adopting a data-driven rather than a rule-driven approach, \tool benefits from the rich expressiveness of real-world code, significantly expanding the search space and enhancing its ability to explore deeper compiler behaviors.

\subsection{Mutation-based compiler testing.} 
Instead of generating a complete program from scratch, another line of research is to mutate parts of an existing test program. Some of the most effective tools maintain semantic equivalence during mutation by leveraging the concept of equivalence modulo inputs (EMI)~\cite{le2014emi}. For instance, Orion~\cite{le2014emi} and Athena~\cite{le2015athena} employ random and guided mutation strategies, respectively, primarily by inserting or deleting dead code blocks. In contrast, Hermes~\cite{sun2016emi} focuses on mutating live code --- sections that are actually executed. Meanwhile, GrayC~\cite{even2023grayc} applies coverage-guided mutations to seed programs, which proves effective in detecting crash bugs, though it has not been successful in uncovering miscompilation bugs.
Beyond semantics-preserving mutations, some approaches modify programs without maintaining their original semantics. For example, classfuzz~\cite{chen2016classfuzz} mutates class files using a diverse set of mutation operations to test JVM implementations. Similarly, LangFuzz~\cite{christian2012langfuzz} adopts a two-phase fuzzing strategy, \ie, learning and mutation, to discover bugs in JavaScript interpreters. It reuses syntax-valid code fragments from seed programs, but performs static, grammar-based substitutions. In contrast, \tool applies LLM-based, context-aware transformations, enabling more expressive and adaptive mutations. 
Negai \etal propose a program generator~\cite{Nagai2014RandomTOv2} for random arithmetic expressions, which employs non-semantics-preserving mutations on the generated expressions. This work builds upon their earlier approach~\cite{Nagai2012RandomTOv1}, which preserved semantics to avoid introducing undefined behavior. In the later work, they enhance the generator by incorporating heuristics to produce more diverse expressions. 

Similar to random test case generators, mutation-based compiler testing is also constrained by predefined mutation rules, which limit the diversity of generated test cases. These approaches often struggle to explore complex or unexpected program behaviors that may trigger deeper compiler bugs.
To overcome this limitation, \tool incorporates iterative synthesis and LLM-based transformations to enhance mutation diversity. Instead of relying solely on fixed mutation rules, our framework leverages real-world programs as a foundation for mutation, ensuring a broader and more expressive test space. By utilizing LLMs, we can intelligently transform code at the syntax level while preserving its structural integrity. 
Through iterative rounds of mutation and synthesis, \tool continuously refines and expands the mutation space, significantly improving its ability to uncover deep-seated compiler bugs.

\subsection{LLM-based Compiler Testing}  
With the rapid advancement of large language models (LLMs), leveraging LLMs for test case generation has emerged as a promising direction in compiler testing.  
Fuzz4All~\cite{xia2024fuzz4all} is the first universal fuzzer that utilizes LLMs as both an input generator and a mutation engine, enabling the testing of widely used systems, including compilers. It employs a large \textit{distillation LLM} to sample multiple candidate prompts, which are then passed to a \textit{generation LLM} to produce diverse test cases.  
WhiteFox~\cite{yang2024whitefox} is the first white-box compiler fuzzer that integrates LLMs with source-code analysis to test deep learning (DL) compiler optimizations. It adopts a multi-agent framework comprising three key components: \textit{Requirement Summarization}, \textit{Test Generation}, and \textit{Feedback Loop}. Given the optimization pass source code from DL compilers, it analyzes the requirements necessary to trigger optimizations, generates corresponding test cases, and feeds valid cases back into the loop for continuous refinement.  

Despite their effectiveness, these existing tools heavily center around LLMs, leading to multiple interactions per test case generation round, which significantly increases computational cost and makes them time-consuming compared to traditional testing approaches.  
In contrast, \tool adopts a hybrid offline/online mode to enhance reuse efficiency, reducing both the computational and financial burden. The generated test cases undergo rigorous validation to ensure high quality, making \tool a more efficient and scalable solution. Evaluation results in Section~\ref{subsec:comparison} demonstrate that \tool outperforms existing LLM-based frameworks.  

\section{Conclusion}
We present \tool, an LLM-based compiler testing framework that decouples the compiler testing process into two synergistic phases: an offline phase and an online phase. In the offline phase, we leverage real-world code-aligned prompting to guide LLMs in generating small, feature-rich, and valid program snippets. These building blocks are then reused in the online phase to synthesize large and complex test programs through iterative program synthesis.
Our evaluation demonstrates the effectiveness of this approach. \tool has uncovered 66 bugs in GCC and LLVM, most of which have already been fixed by compiler developers. Notably, nearly half of the reported bugs are miscompilation bugs, which cannot be detected by previous LLM-based tools.
We believe that \tool opens up a new paradigm of compiler testing, and we are excited to see more research in this direction.

\section*{Data-Availability Statement}
\tool is open-sourced at \url{https://github.com/cuhk-s3/LegoFuzz}. All the source code and data for reproducing the experimental results in this paper are available here~\cite{artifact}.

\bibliographystyle{ACM-Reference-Format}
\input{camera-ready.bbl}

\end{document}